\documentstyle[12pt]{article}
\renewcommand{\thefootnote}{\fnsymbol{footnote}}
\begin{document}
\begin{flushright}
Columbia preprint CU--TP--706
\end{flushright}
\vspace*{1cm}
\setcounter{footnote}{1}
\begin{center}
{\Large\bf The Maximum Lifetime of the Quark--Gluon Plasma\footnote{ This
work was supported by the Director, Office of Energy
Research, Division of Nuclear Physics of the Office of
High Energy and Nuclear Physics of the U.S. Department of
Energy under Contract No.\ DE-FG-02-93ER-40764.}}
\\[1cm]
Dirk H.\ Rischke\footnote{Partially supported by the
Alexander von Humboldt--Stiftung under
the Feodor--Lynen program.} and Miklos Gyulassy \\ ~~ \\
{\small Physics Department, Pupin
Physics Laboratories, Columbia University} \\
{\small 538 W 120th Street, New
York, NY 10027, U.S.A.} \\ ~~ \\ ~~ \\
{\large September 1995}
\\[1cm]
\end{center}
\begin{abstract}
The width $\Delta T$ of the deconfinement transition region
is shown to influence strongly the flow structure in the
(Landau--) hydrodynamical expansion of a quark--gluon plas\-ma.
For a sharp first order transition ($\Delta T=0$)
the mixed phase is rather long-lived, with a lifetime
that has a maximum when the initial energy density is at the phase boundary
bet\-ween mixed and pure quark--gluon matter.
For increasing $\Delta T$, however, the lifetime
decreases rapidly. Hadronic matter, however,
remains long-lived as a consequence of
the rapid change in the degrees of freedom in the transition region
and the corresponding ``softening'' of the equation of state.
\end{abstract}
\newpage
\renewcommand{\thefootnote}{\arabic{footnote}}
\setcounter{footnote}{0}
\section{Introduction}
The hydrodynamical approach has been applied extensively
to describe the dynamics of relativistic heavy--ion collisions
\cite{hydro1,hydro2,hydro3}, because it provides the only
direct link between the equation of state of hot and dense
nuclear matter and observable collective flow phenomena.
Hydrodynamics is defined by local energy--momentum and
(net) charge conservation,
\begin{equation} \label{eom}
\partial_{\mu} T^{\mu \nu} = 0~,~~\partial_{\mu}
{\bf N}^{\mu} =0~.
\end{equation}
$T^{\mu \nu}$ is the energy--momentum tensor, $N_i^{\mu}$
the (net) four--current of the $i$th conserved charge,
${\bf N}^{\mu} = (N_1^{\mu},\, N_2^{\mu},...,\, N_m^{\mu})$ for
$m$ conserved charges.
Under the assumption of local equilibrium (the so-called
``ideal fluid'' approximation) the energy--momentum tensor
and the $m$ (net) charge currents assume the particularly simple
form \cite{LL}
\begin{equation} \label{tmunu}
T^{\mu \nu} = (\epsilon + p)\, u^{\mu} u^{\nu} - p\,
g^{\mu \nu}~,~~N_i^{\mu} = n_i\, u^{\mu}~,~~i=1,...,m~,
\end{equation}
where $\epsilon,\, p,\, n_i$ are energy density, pressure, and
(net) density of the $i$th conserved charge in the
local rest frame of the fluid, $u^{\mu} = \gamma
(1,{\bf v})$ is the four--velocity of the fluid (${\bf v}$ is
the three--velocity, $\gamma \equiv (1-{\bf v}^2)^{-1/2}$,
$u_{\mu} u^{\mu} = 1$), and $g^{\mu \nu} = {\rm diag}(+,-,-,-)$
is the metric tensor\footnote{Our units are $\hbar=c=k_B=1$.}.
The system of $m + 4$ equations (\ref{eom}) is closed via
eliminating one of the $m+5$ unknowns in eqs.\ (\ref{tmunu})
by choosing an equation of state
e.g.\ in the form $p=p(\epsilon,n_1,...,n_m)$.
In the ideal fluid approximation, the (equilibrium) equation of
state is the {\em only\/} input to the hydrodynamical
equations of motion (\ref{eom}) that relates to properties of
the matter under consideration and is thus able to influence
the dynamical evolution of the system. The final results are
uniquely determined once a particular initial condition and a
decoupling (``freeze--out'') hypersurface are specified.

In ultrarelativistic heavy--ion collisions ($\sqrt{s} \sim
200$ AGeV), the central region (in the center-of-momentum
frame) is essentially free of (net) charges, due to the
insufficient amount of baryon stopping \cite{Bj},
and the charge conservation equations in (\ref{eom}) do not
have to be solved explicitly to describe the fluid evolution
in this particular region of space--time.
Moreover, the equation of state depends only
on a single independent thermodynamical variable, for instance
the temperature $T$. From the pressure as a function of $T$,
$p(T)$, one derives the entropy density
$s \equiv {\rm d}p/{\rm d}T$, and the energy density
$\epsilon \equiv Ts-p$.

For this particular case, lattice QCD calculations
\cite{lattice} provide fundamental
information on the equation of state of strongly interacting
matter. These calculations \cite{petersson}
show a rapid increase of the entropy
density as a function of temperature, $s(T)$, in a narrow
temperature interval $\Delta T \sim 10$ MeV
around a temperature $T_c \simeq 150$ MeV,
indicating the presence of a transition from hadronic
degrees of freedom at low temperatures to quark and gluon
degrees of freedom at high temperatures. For a first order phase
transition, the increase in $s$ is actually a discontinuity,
and $\Delta T$ is zero, while for a second or higher
order transition the change of the entropy density is continuous,
and $\Delta T$ is finite. Thusfar, however, lattice calculations
have not converged on the order of the QCD phase transition
\cite{brown,karsch,kanaya}, and we only know that $\Delta T \leq 0.1\, T_c$.

In this paper we discuss the sensitivity of the
structure of hydrodynamic flow to this uncertainty
in the equation of state. To this end
we treat the width $\Delta T$ of the transition region as a free parameter
and assume the following functional form for the
entropy density \cite{blaizot},
\begin{equation} \label{eos}
s(T) = c_H\, T^3\, \frac{1-\tanh [(T-T_c)/\Delta T]}{2}
     + c_Q\, T^3\, \frac{1+\tanh [(T-T_c)/\Delta T]}{2}~.
\end{equation}
Other functional forms are, of course, also possible
and eq.\ (\ref{eos}) is chosen only for the sake of simplicity.
For $T \ll T_c$, the entropy density becomes that of a hadron
gas, $s(T) \simeq c_H T^3$, while for $T \gg T_c$, it assumes
the value for a quark--gluon plasma, $s(T) \simeq c_Q T^3$.
We denote the ratio of degrees of freedom in
the quark--gluon and hadron phases by $r \equiv c_Q/c_H$.
If the hadron phase is taken to be an ideal, massless pion gas, and
the quark--gluon plasma to consist of (non-interacting, massless)
$u$ and $d$ quarks and gluons, then $r=37/3$. In the limit
$\Delta T \rightarrow 0$ the equation of state (\ref{eos})
features a first order phase transition between the two phases
at the phase transition temperature $T_c$ and reduces
to the MIT bag model equation of state \cite{MIT}.
In this paper we vary $\Delta T$ between $0$
and $0.1\, T_c$, consistent with the uncertainties in
present lattice QCD data.

The first study of the sensitivity of hydrodynamic flow on
$\Delta T$ was done in \cite{blaizot}.
It was found that for $\Delta T <  \Delta T^*$,
where $\Delta T^* \simeq 0.07\, T_c$,
matter is thermodynamically anomalous in the phase
transition region and, consequently, that rarefaction
shock waves occur in the one--dimensional expansion
\cite{bugaev,test1}. On the other hand, for $\Delta T \geq
\Delta T^*$, matter is thermodynamically normal everywhere and
simple rarefaction waves form the hydrodynamically stable
solution. The main result of Ref.\ \cite{blaizot}, however, was that,
for the particular initial conditions chosen in that work,
those structural differences had little effect
on the final flow pattern. This led the authors of \cite{blaizot}
to the conclusion that observables,
such as the distribution of outgoing
particles, would hardly be affected by the
uncertainty in $\Delta T$.

Renewed interest in the sensitivity of hydrodynamic
flow on the equation of state has arisen as a
consequence of the work of Refs.\ \cite{test1,hung,flow}.
It was demonstrated that a phase transition
leads to a so-called \cite{hung} ``softest point''
in the equation of state, defined by a minimum in the ratio
$p/\epsilon$. For the equation of state (\ref{eos}) with
$\Delta T=0$, this point occurs at the phase boundary between
mixed and pure quark--gluon matter (see also Section 2).
It was argued \cite{hung} that at this point the
comparatively small pressure prevents a fast expansion and
cooling of the system. This leads to a maximum
in the lifetime of mixed phase matter as a function of
the initial energy density.
On this basis it was suggested \cite{hung}
that the resulting enhanced emission of electromagnetic probes
from this long-lived mixed phase could serve as
a signature for the creation of the quark--gluon plasma
in ultrarelativistic heavy--ion collisions.

In Ref.\ \cite{flow}, the suppressed tendency for expansion
at the softest point was shown to cause a minimum in the
excitation function of the directed transverse collective flow
in semi-peripheral collisions. It was proposed that this
minimum would also be a unique
signature for the phase transition. Another important observable
sensitive to the lifetime of the plasma is pion interferometry
\cite{pratt}. In the standard ``slowly burning log''
scenario inspired by the small deflagration velocities
associated with sharp transitions \cite{vanH} one expects
a dramatic difference between ``out''-- and ``side''--correlations,
which could serve as a signal of the phase transition.

In the present work, we investigate the hydrodynamic flow
pattern for one--dimensio\-nal expansion
in greater detail than in \cite{blaizot}. In particular, we focus on the
question of the lifetime of different parts of the system
as a function of the initial energy density $\epsilon_0$
and the width $\Delta T$ of the transition region.
While one--dimensional hydrodynamics is
not realistic for a detailed comparison with heavy--ion data,
its simplicity allows us to study the influence of the
equation of state on the hydrodynamic flow without additional
effects originating from more complicated geometries.

We confirm the observation made in \cite{blaizot}
that, for initial energy densities well
above or below the softest point of the equation of state,
the temperature profiles in the thermodynamically normal case
($\Delta T \geq \Delta T^*$)
look indeed similar to those in the thermodynamically anomalous
case ($\Delta T < \Delta T^*$).
However, we demonstrate that
the temperature profiles are quite sensitive
to a variation of $\Delta T$ for $\epsilon_0$
{\em near\/} the softest point of the equation of state.
In particular, the lifetime of matter with temperatures
near $T_c$ decreases rapidly for increasing $\Delta T$.
On the other hand, the lifetime of
matter with temperatures well below $T_c$ remains large and may
even grow for increasing $\Delta T$. This is due to the ``softening''
of the equation of state by the change of degrees of
freedom across the transition region.
Thus, the detectability of the quark--gluon plasma via
enhanced emission of electromagnetic probes may depend
sensitively on $\Delta T$, while other observables such as
flow and correlation functions are expected
to remain fairly insensitive to that uncertainty.

This paper is organized as follows. In Section 2 we review
properties of the equation of state (\ref{eos}) in greater
detail and establish the connection to thermodynamically
normal or anomalous behaviour.
Section 3 presents the expansion solutions as
functions of $\epsilon_0$ and $\Delta T$. In Section 4
we discuss the lifetimes of parts of the system having
different temperatures. Section 5 concludes this work with
a summary of our results.

\section{Equation of state and thermodynamical behaviour}

In this section we discuss the equation of state (\ref{eos})
and its thermodynamical properties.
The entropy density at the phase transition temperature is
$s_c \equiv s(T_c) = c_H (r+1) T_c^3/2$.
With the definition $\rho \equiv (r+1)/(r-1)$ and
$\Theta(T) \equiv (T-T_c)/\Delta T$, one can
write \cite{blaizot}
\begin{equation}
\frac{s}{s_c}(T) = \left[\frac{T}{T_c}\right]^3
\left( 1 + \rho^{-1}\, \tanh \Theta(T) \right)~,
\end{equation}
and subsequently obtains pressure and energy density from
\begin{equation} \label{fund}
p (T) = \int_0^T {\rm d}T' \, s(T')~,~~\epsilon (T) =
T\, s (T) - p (T)~.
\end{equation}
The square of the velocity of sound is
\begin{equation} \label{cs2}
c_s^2 \equiv \frac{{\rm d}p}{{\rm d}\epsilon} = \frac{s}{T}
\frac{{\rm d}T}{{\rm d}s} = \left(3 + \frac{T}{\Delta T}~
\frac{1}{\rho + \tanh \Theta(T)}~ \frac{1}{\cosh^2 \Theta(T)}
\right)^{-1}.
\end{equation}
It is obvious that for $T\ll T_c$ or $T\gg T_c$ (or,
equivalently, $|\Theta(T)| \gg 1$) the velocity of
sound approaches the value for an ultrarelativistic ideal
gas, $c_s =1/\sqrt{3}$. In the transition region, $c_s$
decreases and has a minimum.
In the limit $\Delta T \rightarrow 0$,
$c_s$ approaches zero at $T=T_c$ and assumes the ideal gas
value elsewhere.
In Fig.\ 1 we show the energy density,
the pressure, the entropy density, and the square of the
velocity of sound as functions of temperature. Note that above
$T_c$, the energy density approaches the Stefan--Boltzmann value for a
quark--gluon gas from above, while the pressure approaches
it rather slowly from below. Both phenomena are observed in
lattice data \cite{petersson}, and are a consequence of general
thermodynamical relationships \cite{yuki}.

A way to classify the structure of
one--dimensional hydrodynamical solutions
\cite{bugaev,test1} is through the sign of
\begin{equation}
\Sigma \equiv \frac{{\rm d}^2 p}{{\rm d}\epsilon^2} +
\frac{2\, c_s^2 (1-c_s^2)}{\epsilon +p}~.
\end{equation}
(An equivalent classifying quantity is the curvature of the Poisson adiabat
\cite{blaizot}.) For the equation of state (\ref{eos}) we obtain
\begin{equation} \label{SigTs}
\Sigma\, Ts = 3\, c_s^2 \, (1-3\, c_s^2)\left[1 +
\frac{4\, c_s^4}{1-3\, c_s^2}+\frac{2}{3} \, c_s^2\,
 \frac{T}{\Delta T}\, \tanh \Theta(T)\right]~.
\end{equation}
Matter is called thermodynamically normal if $\Sigma$ is
positive and thermodynamically anomalous if it is negative.
It was shown in \cite{blaizot,bugaev,test1} that in the first
case the hydrodynamically stable solution for the
one--dimensional expansion of semi-infinite matter
is a simple rare\-faction wave,
while in the latter case it is a rarefaction shock wave.
Depending on the value of $\Delta T$,
the equation of state (\ref{eos}) features
thermodynamically normal as well as thermodynamically
anomalous regions, as discussed below. The
hydrodynamical solution for the one--dimensional expansion
of semi-infinite matter may thus consist of a sequence of
simple waves and shocks.
For the one--dimensional expansion of finite matter,
the expansion solution is that for semi-infinite matter
as long as the rarefaction waves have not reached the center of
symmetry of the system (usually taken to be the center
of the coordinate system, $x=0$). Afterwards, the solution
no longer consists only of simple waves and shocks
but is more complicated (see also next section).
An analytic solution is so far
only known for thermodynamically normal matter with a
constant velocity of sound \cite{landau}.

To see where matter described by (\ref{eos})
becomes thermodynamically anomalous, it is advantageous to
plot thermodynamic quantities as a function of the energy
density, since in the case $\Delta T=0$, the mapping
$T \rightarrow \epsilon(T)$ is not one-to-one. Moreover,
the beam energy fixes the initial energy density, while
the initial temperature is determined only indirectly through
the equation of state.
In Fig.\ 2 we show the entropy density, the ratio
$p/\epsilon$, the velocity of sound
(squared), and $\Sigma Ts$ as functions of $\epsilon$.
Fig.\ 2 (b) serves to elucidate the notion of a
``softest point'' of an equation of state.
For $\Delta T=0$, $p/\epsilon = 1/3 \equiv c_s^2 = const.$
in the hadronic phase. At the phase boundary bet\-ween
hadronic and mixed phase, $\epsilon_H \equiv
3\, T_c\, s_c/2(r+1) = 0.1125\, T_c\, s_c$ (for
our choice of $r=37/3$), the ratio $p/\epsilon$
starts to decrease, since $p\equiv p_c =
T_c\, s_c/2(r+1) = const.$ in the mixed phase
while $\epsilon$ increases. At the phase boundary
between mixed and quark--gluon matter,
$\epsilon_Q \equiv (4r-1)\, T_c\, s_c/2(r+1)
= 1.8125\, T_c\, s_c$ (for $r=37/3$),
$p/\epsilon$ assumes a minimum which is
the so-called ``softest point'' \cite{hung}.
In the quark--gluon phase, $p/\epsilon$ increases
again and reaches the asymptotic value
$1/3 \equiv c_s^2$ for $\epsilon \rightarrow \infty$.
For increasing $\Delta T$, the softest point is
shifted towards smaller values of $\epsilon$.
In addition, the equation of state becomes
``less soft'' at that point, i.e., the minimum value of
$p/\epsilon$ increases. On the other hand,
there is a ``softening'' at energy densities around $\epsilon_H$.

Figs.\ 2 (c,d) show that, for $\Delta T=0$, the velocity of sound
and $\Sigma$ vanish in the mixed phase, $\epsilon_H \leq
\epsilon \leq \epsilon_Q$. Although matter with a vanishing
$\Sigma$ is not thermodynamically anomalous in a rigorous
sense, this nevertheless suffices to produce a rarefaction shock wave
in the expansion \cite{test1,friman}. For $\Delta T$
being finite, but smaller than a critical value\footnote{This
value was found numerically using eq.\ (\ref{SigTs}).}
$\Delta T^* \simeq 0.07676\, T_c$,
there are always regions where $\Sigma$ becomes negative,
i.e., where the equation of state is thermodynamically
anomalous, and consequently, where rarefaction shock waves
appear in the expansion. For $\Delta T$ larger than
$\Delta T^*$, matter is thermodynamically normal
everywhere and simple rarefaction waves form the
hydrodynamically stable solution. Nevertheless, since
the velocity of sound varies appreciably in the transition
region, Fig.\ 2 (c), there will be modifications to a simple
wave profile calculated with a constant velocity of sound.

Note especially in Fig.\ 2 (d) that the region where matter
becomes thermodynamically anomalous shrinks drastically even
for tiny $\Delta T$. To understand this we first
expand the velocity of sound
squared (\ref{cs2}) for small $\Delta T/T_c \ll 1$
in the vicinity of $T=T_c$,
\begin{equation} \label{cs2appr}
c_s^2 \simeq \frac{\Delta T/T_c}{T/T_c}~\frac{\rho + \tanh
\Theta(T)}{1-\tanh^2 \Theta(T)}~.
\end{equation}
Note that $\Theta(T)\equiv (T-T_c)/\Delta T$ is
{\em not\/} necessarily small in this
limit, i.e., one {\em must not\/} expand the hyperbolic tangents in this
expression. In accord with Figs.\ 1 (d) and 2 (c),
$c_s^2$ is of the order $\Delta T/T_c \ll 1$ near $T_c$.
Next, we determine the roots of eq.\ (\ref{SigTs})
to lowest order in the small quantities $\Delta T/T_c$ and
$T/T_c -1$. This yields a quadratic equation for $\tanh
\Theta (T)$ with only one physical solution $\tanh \Theta(T_{\Sigma})
= \rho - (\rho^2+3)^{1/2} \simeq - 0.9173$, corresponding to the upper
boundary of the region where matter is thermodynamically anomalous. The lower
boundary cannot be obtained in our approximation, because
$c_s^2$ is already large in the respective region of
energy densities, cf.\ Figs.\ 2 (c,d). Since $\tanh \Theta(T_{\Sigma})$
is of order unity, an expansion of the hyperbolic
tangent in (\ref{cs2appr}) is indeed not possible.
Note that $T_{\Sigma}/T_c-1 \equiv (\Delta T/T_c)\, {\rm Artanh} [\rho
- (\rho^2+3)^{1/2}] \simeq -1.572\, \Delta T/T_c$. Thus,
matter with tempe\-ra\-ture $T=T_c$ is always
thermodynamically normal for finite $\Delta T$.
Finally, we calculate the energy density corresponding
to the upper boundary to lowest order,
\begin{eqnarray}
\epsilon_{\Sigma} & \simeq & T_c\, s_c\, (1+\rho^{-1}\, \tanh
\Theta(T_{\Sigma})) - p_c
= \epsilon_c + T_c\, s_c \left(1- \sqrt{1+3\, \rho^{-2}}\right) \nonumber \\
& = & 0.18275\, T_c\, s_c \ll \epsilon_Q = 1.8125\, T_c\, s_c~,
\label{ecrit}
\end{eqnarray}
where $\epsilon_c \equiv T_c\, s_c - p_c = (2r+1)\, T_c\, s_c/2(r+1)
= 0.9625\, T_c\, s_c$. (This is
the point where all curves in Fig.\ 1 (a) intersect.)
This result is remarkable in several aspects. First, the fact that
$\tanh \Theta(T_{\Sigma})$ is not small leads to a large
shift of $\epsilon_{\Sigma}$ away from the point $\epsilon_c$ {\em already
to zeroth order in\/} $\Delta T/T_c$ .
We checked numerically that first order corrections in $\Delta T/T_c$ do
not change the value for $\epsilon_{\Sigma}$ appreciably
before our approximations break down anyway.
Second, this shift is {\em negative}, i.e., away from
$\epsilon_Q$ which is the upper boundary in the case $\Delta T=0$.
Thus, the size of the region where matter is thermodynamically
anomalous does not change continuously in the limit
$\Delta T \rightarrow 0$, but increases abruptly
when going from arbitrarily small $\Delta T$
to $\Delta T=0$. In this sense, the case $\Delta T\equiv 0$ is very
special in that it features an exceptionally large region where
matter is thermodynamically anomalous.

\section{One--dimensional expansion solutions}

In this section the one--dimensional
hydrodynamic expansion of a finite slab of matter
with (homogeneously distributed) initial energy density
$\epsilon_0$ is investigated with emphasis on
the dependence of the flow structure
on $\epsilon_0$ and $\Delta T$.
To solve the hydrodynamic equations (\ref{eom}) in
(1+1)--dimensions we employ the relativistic
Harten--Lax--van Leer--Einfeldt (HLLE) algorithm presented
in \cite{test1,schneider} with a grid spacing
of $\Delta x = 0.025\, R$ ($2\, R$ being the initial size of
the slab) and a time step width of $\Delta t =0.99\, \Delta x$.
This algorithm was shown \cite{test1}
to accurately reproduce hydrodynamic flow profiles for
analytically solvable test problems where the equation of
state has thermodynamically normal as well as
anomalous regions. In particular, for the choice
$\Delta x=0.025\, R$, the algorithm accurately reproduced
the expansion solution for a finite system of size $2\, R$
consisting of thermodynamically normal matter with
a constant velocity of sound. (This solution
was first provided by Landau in purely analytical
form \cite{landau}, but can also be easily constructed
semi-analytically via the method of characteristics
\cite{baym}.)

In the following Figs.\ 3 -- 7 we demonstrate how the flow structure changes
systematically in the range $\epsilon_H \leq \epsilon_0 \leq 10\, \epsilon_Q$
and for various $\Delta T$. Each figure corresponds
to a fixed $\epsilon_0$, and shows the time evolution of temperature,
center-of-momentum frame (CM) energy density, and the
space--time structure of isotherms\footnote{For $\Delta T=0$
all states of mixed phase matter have the same
temperature $T_c$, cf.\ Fig.\ 1 (a).
In this case, we define the $(T=T_c)$--isotherm
as the set of space--time points where $T$ has just
infinitesimally dropped below $T_c$ \cite{test1}.} in
three columns corresponding to
$\Delta T=0,\, 0.01\, T_c,$ and $0.1\, T_c$.

\subsection{$\epsilon_0 = \epsilon_H$}

In Fig.\ 3 we show the expansion solution for
an initial energy density $\epsilon_0 \equiv
\epsilon_H=3\, T_c\, s_c/2(r+1)=0.1125\, T_c\, s_c$.
For $\Delta T=0$, Figs.\ 3 (a,d,g), this energy density
corresponds to the boundary between hadronic and mixed
phase matter.
Thus, thermodynamically anomalous regions of the equation
of state are not reached during the expansion, and the expansion
proceeds at first as a simple rarefaction wave. After this
simple wave overlaps with its symmetric counterpart coming from
the $(-x)$--direction, the hydrodynamical solution is identical
to that given by Landau \cite{landau}.

For $\Delta T=0.01\, T_c$, Figs.\ 3 (b,e,h), the expansion
seems to proceed rather similar to the previous case.
Fig.\ 2 (d) shows, however, that around $\epsilon_0 =
\epsilon_H$ the equation of state has a region where
matter is thermodynamically anomalous. Thus, a rarefaction
shock wave is bound to form in the expansion.
Since the anomalous region is confined to a very narrow
range about $\epsilon_H$, cf.\ Fig.\ 2 (d),
the shock is weak and difficult to resolve numerically.
The best evidence for its existence can be
found in Fig.\ 3 (e), in that the slope of the rarefaction
wave travelling into the slab is steeper than in the
corresponding Figs.\ 3 (d,f) where no shock occurs.

For $\Delta T=0.1\, T_c$, Figs.\ 3 (c,f,i),
matter is thermodynamically normal everywhere,
cf.\ Fig.\ 2 (d). Consequently, the expansion
initially proceeds as a simple rarefaction
wave, as for $\Delta T=0$. For such a wave, the matter
velocity with respect to the wave profile equals the velocity
of sound \cite{LL,test1}. Thus, the head of the simple wave
travels with the velocity of sound $c_s$ into matter which is
initially at rest.
In the respective range of energy densities, $c_s$
is considerably smaller for $\Delta T=0.1\, T_c$ than
for $\Delta T=0$, cf.\ Fig.\ 2 (c). Therefore,
as can be clearly seen in Figs.\ 3 (c,f), it takes much
longer for the simple rarefaction wave to reach the center
$x=0$ than in Figs.\ 3 (a,d).
The delayed rarefaction can be also seen comparing
Figs.\ 3 (g) and (i): it takes about $4\, R$ for the
system at $x=0$ to cool below $T=0.7\, T_c$ and about
$6\, R$ to cool below $T=0.5\, T_c$,
while the corresponding values
for $\Delta T=0$ are $3\, R$ and $5\, R$.

The reason why there is no $(T=T_c)$--isotherm in Fig.\ 3 (h)
and no $(T=T_c)$-- and $(T=0.9\, T_c)$--isotherm in Fig.\ 3 (i)
can be inferred from Fig.\ 1 (a): as long as
the energy density is below $\epsilon_c \equiv
(2r+1)\, T_c\, s_c/2(r+1) = 0.9625\, T_c\, s_c$,
the temperature corresponding to that energy density is
smaller for larger $\Delta T$.
Thus, for $\Delta T=0.01\, T_c$, already the initial
temperature is below $T_c$, and for $\Delta T=0.1\, T_c$
even below $0.9\, T_c$, cf.\ Figs.\ 3 (b,c). Therefore,
there simply are no corresponding isotherms in the space--time
diagram.

Note that the temperature and CM energy density profiles show a
minimum at $x=0$ for late times $t$.
As was explained in \cite{test1}, this is essentially
a relativistic effect. At finite $x$, matter is moving,
while due to symmetry it is at rest at $x=0$. Moving matter,
however, experiences relativistic time dilation and is
thus less diluted and still hotter at a given time
in the CM frame than matter at rest.
Of course, the degree of dilation a fluid element experiences
is determined by its velocity. This velocity is, in turn,
determined by the ``stiffness'' of the equation of state, i.e.,
the ratio of pressure to energy density, since
that ratio influences the acceleration of fluid elements.
Thus, for the typical energy densities
occurring in Fig.\ 3, matter velocities are smaller
for larger values of $\Delta T$, since the
equation of state is softer, cf.\ Fig.\ 2 (b).
This explains why the minimum at $x=0$ (or the time dilation
effect, respectively) is less
pronounced (and occurs at later times)
in Figs.\ 3 (c,f) than for instance in Figs.\ 3 (a,d).

\subsection{$\epsilon_H < \epsilon_0 < \epsilon_Q$}

Fig.\ 4 shows the time evolution of temperature and CM energy
density, and the isotherms in the space--time diagram
for an initial energy density $\epsilon_0= 30\, T_c\, s_c/
2(r+1) = 1.125\, T_c\, s_c$, i.e., $\epsilon_H < \epsilon_0 <
\epsilon_Q = (4r-1)T_c\, s_c/2(r+1)=1.8125\, T_c\, s_c$.
For $\Delta T=0$, Figs.\ 4 (a,d,g),
the initial state is thermodynamically anomalous mixed phase
matter. Therefore, rarefaction proceeds via a shock wave,
in front of which mixed phase matter is at rest.
Hadronic matter emerges from the shock in an accelerated state
and is subsequently diluted by a simple rarefaction wave
\cite{test1}. Note that the shock velocity is quite small,
leading to a long lifetime of $\sim 26\, R$ for mixed phase
matter at $x=0$. We can understand the value for the lifetime from
the expression for the rarefaction shock velocity,
eq.\ (70) of Ref.\ \cite{test1}, which reduces, for
$\epsilon_H \ll \epsilon_0 \leq \epsilon_Q$, to
\begin{equation} \label{appr}
v_{sh} \simeq - \frac{2}{3 \sqrt{3} \, \epsilon_0/\epsilon_H} \simeq -
(2.6\, \epsilon_0/\epsilon_H)^{-1}~.
\end{equation}
Therefore, we expect for $\epsilon_0/\epsilon_H =
2(r+1) \epsilon_0/ 3\, T_c\, s_c = 10$ a lifetime
$t_{life} \equiv R/|v_{sh}| \simeq 26\, R$, in accord with Fig.\
4 (g).

Again, the expansion for $\Delta T=0.01\, T_c$ looks similar,
especially the temperature profiles, cf.\ Figs.\ 4 (a) and (b).
The difference is that now the initial energy density is
well above the region where matter is thermodynamically
anomalous, cf.\ Fig.\ 2 (d). Therefore,
the expansion has at first to proceed as a simple rarefaction
wave, until the energy density has dropped sufficiently to
reach that region. Then, a rarefaction shock wave
provides further dilution. This can be most clearly seen in
the CM energy density profiles, Fig.\ 4 (e). The reason why
the temperature profiles, Fig.\ 4 (b), do not clearly exhibit
this behaviour is that $T$ is proportional to a fractional
power of $\epsilon$, so that
the range of high energy densities is not well resolved when
plotted as a function of temperature. A clear indication for
this behaviour, however, can be found comparing
the space--time diagram Fig.\ 4 (g), where the $(T=0.9\, T_c)$--
and $(T=T_c)$--isotherms coincide, since both temperatures
occur at the position of the shock, with Fig.\ 4 (h),
where they are well separated due to the simple rarefaction
wave preceding the shock. Note that this
has a drastic influence on the lifetime of
matter with $T=T_c$. Such matter is thermodynamically
normal for finite $\Delta T$, as discussed in the context of Fig.\
2 (d), and consequently rapidly diluted by the simple rarefaction wave.
The lifetime drops by more than a factor of 2
as compared to Fig.\ 4 (g). On the other hand, the lifetime of matter with
$T = 0.9\, T_c$ remains comparable to that in the case
$\Delta T=0$, since this temperature occurs on the rarefaction shock front,
cf.\ Fig.\ 4 (b), as was also the case in Fig.\ 4 (a).
We will discuss this phenomenon in more detail in the next section.

For $\Delta T=0.1\, T_c$, Figs.\ 4 (c,f,i), matter is
thermodynamically normal everywhere and the hydrodynamically stable
rarefaction solution is a simple wave (until it reaches
the center $x=0$ and overlaps with its
counterpart from the $(-x)$--region).
Nevertheless, due to the variation of the velocity of sound in
the phase transition region, cf.\ Fig.\ 2 (c), the
shape of the simple wave exhibits a ``bump'' faintly
resembling a smeared version of the rarefaction shock
occurring in the previous two cases.
This bump, however, is moving rapidly in positive
$x$--direction, in contrast to the rarefaction shock in
Fig.\ 4 (b), which is approximately stationary,
or that in Fig.\ 4 (a), which is even slowly moving in
$(-x)$--direction. This has the effect that the hot parts of
the system ($T \geq 0.9\, T_c$) cool faster than in the
two previous cases. This can be understood
with the help of Fig.\ 2 (b) which shows that for the respective
energy densities the equation of state is ``stiffer''
for increasing $\Delta T$, thus increasing
the system's tendency to expand and cool.

On the other hand, the cooler parts ($T \leq 0.7\, T_c$)
keep their temperature longer for $\Delta T=0.1\, T_c$,
cf.\ also Figs.\ 4 (g,h,i). The reason is that the
equation of state for small energy densities (low temperatures)
is softer than in the other two cases, cf.\ Fig.\ 2 (b).
Thus, the acceleration and consequently the matter velocities
are smaller, and therefore the rarefaction and cooling take longer.
We finally mention that, since now $\epsilon_0 > \epsilon_c$, the
initial temperature increases with $\Delta T$,
cf.\ Fig.\ 1 (a), in contrast to the previous case, Fig.\ 3.

\subsection{The softest point $\epsilon_0 \sim \epsilon_Q$}

Fig.\ 5 shows the expansion solutions for
$\epsilon_0=50\, T_c\, s_c /2(r+1) = 1.875\, T_c\, s_c$.
For $\Delta T =0$, Figs.\ 5 (a,d,g), this energy density is
just above the softest point of the equation of state, i.e., the
phase boundary between mixed and pure quark--gluon matter,
$\epsilon_Q \equiv (4r-1)T_c\, s_c/2(r+1) = 1.8125\, T_c\, s_c$.
Consequently, a simple rarefaction wave
precedes the rarefaction shock wave travelling into
hot quark--gluon matter \cite{test1}.
(The simple rarefaction wave is not visible in Fig.\ 5 (a)
due to the large time step chosen.)
This has the consequence that matter is pre-accelerated
before it enters the shock wave. In conjunction with
the equation of state becoming increasingly stiffer for smaller
energy densities in the mixed phase, cf.\ Fig.\ 2 (b),
this has the further consequence that the CM energy
density develops a minimum at $x=0$ for given $t$,
cf.\ Fig.\ 5 (d), just as in Fig.\ 3.
We note that, in agreement with the results
of \cite{test1}, the rarefaction shock is almost stationary
for an $\epsilon_0$ this close to the softest point of the
equation of state. This leads to a (local) maximum for the
lifetime of matter with $T=T_c$ at $x=0$. The particular value
in Fig.\ 5 (g) can be again
understood with the help of the approximate formula
(\ref{appr}) which gives\footnote{We may safely neglect the
small pre-acceleration of matter due to the simple rarefaction
wave preceding the shock in the quark--gluon phase.}
$t_{life} \simeq R/|v_{sh}| \simeq 43\, R$.

The temperature profiles for $\Delta T=0.01\, T_c$, Fig.\ 5 (b),
look again similar to that of Fig.\ 5 (a), although the shock
wave is no longer stationary, but at first
advancing in $x$--direction.
The reason is that the shock strength is smaller
than in the previous case (due to the much narrower
region where matter is thermodynamically anomalous, cf.\ Fig.\ 2 (d)),
which is especially obvious from
comparing Figs.\ 5 (d) and (e). Therefore, the velocity of
the shock, which is originally directed inwards, is too small
to compete with the pre-acceleration of matter
in the simple rarefaction wave preceding the shock,
resulting in a net outward motion of the shock.
This mechanism is also responsible for
the positive shock velocity observed for higher $\epsilon_0$
in the cases $\Delta T=0$ and $0.01\, T_c$, see Figs.\ 6, 7 below.
For later times, however, just before the high energy density
matter in the center is completely consumed, the shock
reverses its direction, due to the fact that the energy density
(or pressure) in the interior has now become too small to keep
it driving outwards. (This effect is also visible in Fig.\ 5 (a).)

The energy density profiles, Fig.\ 5 (e), do not show
the minimum at $x=0$ typical for relativistic time dilation
of moving matter. The reason is that in this case
the equation of state is soft for the respective
energy densities, cf.\ Fig.\ 2 (b). As explained above,
this leads to small (i.e.\ non-relativistic) matter
velocities and the absence of an appreciable amount of time dilation.

The expansion for $\Delta T=0.1\, T_c$, Fig.\ 5 (c,f) resembles the
corresponding one in Figs.\ 4 (c,f). In particular, the
``bump'' structure caused by the variation of the
velocity of sound is again clearly visible. As in Fig.\ 4, however, the
expansion differs appreciably from the one for
$\Delta T=0$ and $0.01\, T_c$ in Figs.\ 5 (a,b,d,e).

The relative time scales of cooling are similar to the
previous case $\epsilon_0 = 1.125\, T_c\, s_c$, cf.\
Figs.\ 4 (g,h,i) with Figs.\ 5 (g,h,i).  Note again the
sharp decrease in the lifetime of matter with $T=T_c$ for
$\Delta T=0.01\, T_c$ as compared to the case $\Delta T=0$,
while matter with $T=0.9\, T_c$ remains long-lived. For
$\Delta T=0.1\, T_c$, also the lifetime of matter with
temperature $ T = 0.9\, T_c$ is reduced, while
matter that is even cooler still lives longer.

\subsection{$\epsilon_0 \sim 2\, \epsilon_Q$}

Fig.\ 6 shows the expansion for an initial energy density
$\epsilon_0 = 100\, T_c\, s_c/2(r+1) = 3.75\, T_c\, s_c
\sim 2 \, \epsilon_Q$.
This case is of interest since for initial energy
densities of this order of magnitude a rather peculiar
phenomenon occurs in the expansion for an equation of state
with $\Delta T=0$. One observes in Figs.\ 6 (a,d) that the
rarefaction shock wave is now driven outwards by the
pre-acceleration from the simple rarefaction wave, as was
already discussed above. Matter behind the shock
experiences cooling typical for the Landau expansion with
a sufficiently stiff equation of state, i.e., slower
matter near the center $x=0$ cools more rapidly than faster
matter at larger $x=0$. This, however, has the consequence
that as soon as the (comoving) energy density drops below
$\epsilon_H$, or equivalently, the temperature drops below
$T_c$, a second rarefaction shock wave develops
that connects hadronic matter in the center with
mixed phase matter behind the first shock.

The occurrence of the second shock could have been expected,
since matter changes from thermodynamically anomalous in
the mixed phase to thermodynamically normal in the
hadron phase. To our knowledge there is now,
however, no general mathematical theorem
rigorously proving the existence of this second shock,
as was the case in the expansion of semi-infinite
matter studied in \cite{test1}. The shock has an
interesting further consequence: just as the first shock
expells hadronic matter outwards, the second accelerates
matter {\em inwards\/}. Therefore, there is
an accumulation of matter in the center which is consequently
{\em reheated\/} instead of cooled. This reheating of the
center can be clearly seen in Fig.\ 6 (g), where the
triangle-shaped isotherm at small $x$ and $t\sim 70\, R$ is a
{\em re-occurrence\/} of the $(T=0.9\, T_c)$--isotherm caused
by reheating. In addition, also the $(T=0.7\, T_c)$--isotherm
shows that the central region remains hot over a much longer
period of time than for the expansion of matter without a
phase transition.

The reheating is considerably smaller for $\Delta T = 0.01\, T_c$,
cf.\ Figs.\ 6 (b,e,h), and basically only discernible
from the shape of the $(T=0.7\, T_c)$--isotherm. The reason
is, of course, that the region in the equation of state,
where matter is thermodynamically anomalous, is much narrower, cf.\ Fig.\
2 (d), so that the strength of the second rarefaction shock
is smaller and consequently the acceleration of matter
towards the center much weaker. For $\Delta T=0.1\, T_c$
there is no such an effect at all, and the fluid evolution is
rather different from the cases $\Delta T=0$ and $0.01\, T_c$.

\subsection{$\epsilon_0 \gg \epsilon_Q$}

Finally, we show in Fig.\ 7 the expansion for $\epsilon_0
= 500\, T_c\, s_c /2(r+1)= 18.75\, T_c\, s_c \gg \epsilon_Q$.
This case is similar to the one studied in Fig.\ 13 in Ref.\
\cite{blaizot}, although we followed the fluid
evolution over a much longer time. Here,
the acceleration from the simple wave travelling into the
quark--gluon phase is rather large, leading to a
shock wave in Figs.\ 7 (a,b,d,e) that is rapidly
moving outwards. Note that now also the temperature profiles for
$\Delta T=0.1\, T_c$, Fig.\ 7 (c), resemble
those in the case $\Delta T=0$, Fig.\ 7 (a)
(with the exception of the immediate vicinity
of the rarefaction shock).
The main reason is that the fluid motion is now dominated
by the conversion of the large initial energy density into
collective motion, so that the influence of the phase
transition region on the structure of the hydrodynamic
solution becomes less important. Moreover, as in the cases
considered previously, the variation of the velocity of sound in
the phase transition region causes a ``bump'' structure that
resembles a smeared shock.

The similarity of the temperature profiles in Figs.\ 7 (a)
and (c) led the authors of \cite{blaizot}
to the conclusion that final state observables will not be
much influenced by the precise value of $\Delta T$. They reached the
same conclusion studying the expansion with an initial entropy density
$s_0 = 0.3\, s_c$ (Fig.\ 11 in \cite{blaizot}) which corresponds, for
$\Delta T=0$, to an initial energy density $\epsilon_0 = 0.2625\,
T_c\, s_c$. For this case, the expansion is similar to the one studied here
in Fig.\ 3. That figure shows that the influence of $\Delta T$ on
the expansion is negligible also for such small initial energy densities,
confirming the conclusions of \cite{blaizot}.
As we have seen, however, the profiles are {\em rather\/} sensitive to the
choice of $\Delta T$ for initial energy densities
{\em near the softest point of the equation of state\/}. In particular,
the reheating phenomenon of Figs.\ 6 (a,d,g)
does not occur at all for $\Delta T=0.1\, T_c$, cf.\
Figs.\ 6 (c,f,i). We discuss the effect on the lifetime of matter
with different temperatures in the next section.

\section{Lifetime of parts of the system having different temperatures}

In this section we discuss the lifetimes of parts of the system
having different temperatures. The ``lifetime'' is defined as
the time it takes the system to cool below a
certain temperature $T^*$ at $x=0$. More precisely, the ``lifetime'' is the
intercept of the isotherm $t(x;T^*)$ in the $(t-x)$--plane
with the $t$--axis, i.e., $t_{life} \equiv t(0;T^*)$.
We choose the point $x=0$ to measure this time, since the
motion of fluid cells at finite $x$ causes a time dilation
which leads to a slower cooling when measured
in the global CM frame of the system.
In case there is more than one intercept, for instance when
there is reheating of the center, Fig.\ 6 (g), we choose
the first.

We investigate the lifetime in two different geometries. In
subsection 4.1 we study longitudinal expansion for fixed initial
``radius'' $R$ of the system.
In subsection 4.2 we investigate a more realistic geometry
where the initial longitudinal size $R$ takes
into account the Lorentz contraction of the colliding nuclei.
Thus, $R$ varies with beam energy, and hence with the initial
energy density $\epsilon_0$.

\subsection{Fixed $R$}

In Fig.\ 8 we show the lifetime of parts of the system having
temperature $T=0.5\, T_c$,
$0.7\, T_c,\, 0.9\, T_c,\, T_c,\, 1.1\, T_c,$
and $1.3\, T_c$ for the three different
$\Delta T=0,\, 0.01\, T_c,\, 0.1\, T_c$, and for an
ideal gas equation of state without phase transition and
constant velocity of sound $c_s=1/\sqrt{3}$
as a function of the initial energy density $\epsilon_0$.
We measure the lifetime in units of $R$.
For the ideal gas calculation, Fig.\ 8 (d), there
is of course no phase transition, and the temperature
scale set by $T_c$ is arbitrary. The
results of Fig.\ 8 (d) were obtained using
$T_0/T_c \equiv (4\, \epsilon_0/3\, T_c\, s_c)^{1/4}$.

One observes that for initial energy
densities smaller than $\sim \epsilon_c$,
the lifetime of matter with temperatures $T \leq 0.9\, T_c$ is
essentially independent of $\Delta T$. The only exception is
that for $\Delta T=0$ and $0.01\, T_c$ the lifetime of matter
having $T=0.7\, T_c$ and $T=0.9\, T_c$ is equal,
while that is not the case for $\Delta T=0.1\, T_c$.
The reason is obvious from Figs.\ 4, 5:
in the first two cases the rarefaction shock wave, once it
reaches the center, abruptly cools matter to about $0.5\, T_c$,
thus terminating the existence of matter having $T=0.9\, T_c$
as well as $T=0.7\, T_c$ simultaneously.
This can be clearly seen in the space--time
diagrams Figs.\ 4, 5 (g) and (h).
In the case $\Delta T = 0.1\, T_c$,
rarefaction proceeds via simple waves
which are continuous solutions and
gradually cool the system, and thus matter with
$T=0.7\, T_c$ survives longer than hotter matter with
$T=0.9\, T_c$.

The second observation is that the lifetime of
``very hot'' parts of the system, i.e., matter having
temperature $T=1.1\, T_c$ and $1.3\, T_c$,
is also comparable for all three
values of $\Delta T$. These high temperatures occur,
of course, only for sufficiently high initial energy
densities. The reason for the similarity in the lifetimes
is that a variation of $\Delta T$ has little effect on the
equation of state for high energy densities (or temperatures), cf.\
Fig.\ 1. Consequently, also the fluid evolution must be similar
for these high temperatures .

For initial energy densities near the softest point of the
equation of state and for temperatures around $T_c$,
however, $\Delta T$ has an appreciable
influence on the lifetime.
For instance, one clearly observes in Fig.\ 8 (a)
that the lifetime of matter with $T=0.9\, T_c$ and
$T= T_c$ has a maximum near the softest point,
cf.\ Fig.\ 2 (b). This effect was discussed in detail in
Ref.\ \cite{test1} and in the context of Fig.\ 5 above. It
has been explained by the existence of
a minimum in the velocity of the rarefaction wave travelling
into the system, leading to the ``slow burning'' of
hot matter also observable in Figs.\ 4, 5 (a,d).
An alternative explanation was given in \cite{hung},
namely that the small $p/\epsilon$--ratio at this
point reduces the system's tendency to expand.

However, Figs.\ 8 (b,c) show that this maximum in the lifetime
is rapidly washed out and eventually vanishes for increasing $\Delta T$,
although the $p/\epsilon$--ratio still exhibits a minimum,
cf.\ Fig.\ 2 (b). This vanishing of the maximum in the lifetime
of matter with a particular temperature is essentially explained
by a gradual {\em reduction\/} of the lifetime when
increasing $\Delta T$: first,
the lifetime of matter with $T=T_c$ is reduced when
increasing $\Delta T$ from $0$ to $0.01\, T_c$, cf.\ Figs.\
8 (a) and (b). Then, that of matter with $T=0.9\, T_c$ is reduced
when further increasing $\Delta T$ from $0.01\, T_c$ to $0.1\, T_c$, cf.\
Figs.\ 8 (b) and (c). The lifetime of cooler matter with $T\leq 0.7\, T_c$
is not appreciably influenced around $\epsilon_0 = \epsilon_Q$.

In order to understand this reduction of the lifetime for
increasing $\Delta T$, we show in Fig.\ 9 the
time evolution of the temperature at $x=0$ for various values of
$\Delta T$ and an initial energy density $\epsilon = 50\, T_c \,
s_c/ 2(r+1) = 1.875\, T_c \, s_c$. This value is close to
the softest point of the equation of state, i.e., where the above
mentioned effects on the lifetime are most drastic. In this
representation, the rather long lifetime of matter with temperature $T=T_c$
($t_{life} \simeq R/|v_{sh}| \simeq 43\, R$)
for a sharp first order phase transition can be seen in a particularly
clear way. This {\em temporal\/} temperature profile changes
{\em smoothly\/} with increasing $\Delta T$. The {\em drastic\/}
effect on the lifetime of matter with $T=T_c$ and $T=0.9\, T_c$
observed in Fig.\ 8 is readily explained via Fig.\ 9 as follows.
For $\Delta T=0$
the temperature stays constant (and equal to $T_c$) over a rather long period
of time $\sim 43\, R$, due to the small velocity of the rarefaction shock
which consumes mixed phase matter and consequently takes a long
time to reach the center $x=0$, cf.\ Fig.\ 5 (a). After that, the
temperature drops sharply, cf.\ also Fig.\ 5 (g).

For small finite $\Delta T = 0.01\, T_c$, however,
the temperature is {\em already slowly decreasing
with time}, cf.\ also Figs.\ 5 (b,e).
This leads to a much shorter lifetime of matter with the particular
temperature $T=T_c$, although the system remains obviously
rather hot, $T\simeq T_c$, for a time span $\sim 37\, R$ that is
only slightly smaller than in the previous case.
This also accounts for the long lifetime of matter with $T=0.9\, T_c$ as
observed in Fig.\ 8 (b). After that time, the cooling is again
rather drastic due to the rarefaction
shock which has then reached the center $x=0$.
For increasing $\Delta T$, the cooling of hot matter becomes
faster, which explains why the lifetime of matter with $T=0.9\, T_c$
is eventually also appreciably reduced, cf.\ Fig.\ 8 (c).

One also notes that the lifetime of matter with $T=0.7\, T_c$ is not
much influenced when increasing $\Delta T$, in accord with our
above observation in Fig.\ 8. On the other hand,
matter with even lower temperatures may survive longer.
In general, the more rapid cooling of matter with
higher temperatures as well as the delayed cooling of matter
with smaller temperatures is well explained by the increasing
``stiffness'' of the equation of state at higher energy densities
and its ``softening'' at smaller energy densities for increasing $\Delta T$,
cf.\ Fig.\ 2 (b).

An important conclusion to be drawn from Figs.\ 8 and 9 is that
for increasing $\Delta T$ the maximum in the
lifetime of matter with temperature around $T_c$ vanishes
and that, for $\Delta T= 0.1\, T_c$,
there is no longer any irregularity that can be traced to the
rapid variation of thermodynamical quantities in the
phase transition region. In fact, in that case the shapes of
the lifetime curves resemble those for
the expansion of an ideal gas, cf.\ Figs.\ 8 (c) and (d). However,
the existence of a transition from hadronic to quark and gluon
degrees of freedom, the softening of the equation of state induced by that
transition, and the resulting delayed expansion
causes the lifetimes to be larger by an overall factor
$\sim 10$, cf.\ also Fig.\ 9. We will discuss possible implications
for proposed quark--gluon--plasma signatures in the next section.

The curves in Fig.\ 8 (d) could also be calculated analytically
employing Landau's solution \cite{landau} which gives
\begin{equation}
t|_{x=0} = \left. \frac{\partial \chi(T,\alpha)}{\partial T}
\right|_{x=0}~,
\end{equation}
where
\begin{equation}
\chi(T,\alpha) = - \frac{T R}{c_s} \int_{c_s \alpha}^{\ln [T_0/T]}\,
{\rm d}u~e^{(1+\beta)u}~I_0\left(\beta \sqrt{u^2-c_s^2 \alpha^2}\right)~.
\end{equation}
Here $\beta \equiv (1-c_s^2)/2c_s^2 =1$ for $c_s^2=1/3$
and $\alpha \equiv \tanh v$ is the fluid rapidity.
At $x=0$ we also have $\alpha=0$ by symmetry and we obtain
\begin{equation} \label{t1}
t|_{x=0} = \frac{R}{c_s} \left\{ \left( \frac{T_0}{T} \right)^2
I_0 \left(\ln [T_0/T] \right) - \int_0^{\ln [T_0/T]}\,
{\rm d}u~e^{2u}~I_0(u) \right\}~.
\end{equation}
To understand the qualitative behaviour of the curves in Fig.\ 8 (d),
it is sufficient to note that the integral is always dominated by
contributions from the upper boundary.
For $T_0 \gg T$ one may thus use the asymptotic approximation
$I_0(u) \sim e^u$ in the integrand and obtains
\begin{equation} \label{t2}
t|_{x=0} \equiv t_{life} \sim R\, T_0^3 \sim R\, \epsilon_0^{3/4}
\end{equation}
for fixed $T$. Thus, $t_{life}/R$ behaves (at least asymptotically)
$\sim \epsilon_0^{3/4}$ {\em independent\/} of $T/T_c$. This behaviour
is in accord with the one observed in Fig.\ 8 (d). Curves for
different $T/T_c$ are simply shifted due to different constants
of proportionality. Note that an increase of the lifetime
proportional to $\epsilon_0^{3/4}$ is also expected in
scaling hydrodynamics \cite{test1}.

\subsection{Variable $R(\epsilon_0)$}

In Fig.\ 8  $R$ was assumed fixed for all initial energy densities.
Now we allow it to vary with $\epsilon_0$ to take
into account the Lorentz contraction of the colliding nuclei.
To specify this relationship quantitatively, we consider
two models for the initial stage of central $A+A$--collisions.

The first is the scenario employed in \cite{hung}. It is assumed
that half of the available CM energy, $\sqrt{s}=2\, A M_N \gamma_{CM}$,
is deposited in a cylinder of height $2\, R_A/\gamma_{CM}$
and radius $R_A$. Thus, $\epsilon_0 = A M_N \gamma_{CM}^2 / 2 \pi
R_A^3$. The initial ``radius'' of the system is therefore
\begin{equation} \label{ed}
R=\frac{R_A}{\gamma_{CM}}= \frac{R_A}{(\epsilon_0/T_c\, s_c)^{1/2}}
\left(\frac{A M_N}{4 \pi (r+1)p_c R_A^3}\right)^{1/2}~.
\end{equation}
With $p_c = \pi^2 T_c^4/30$, $T_c = 170$ MeV \footnote{This
value is slightly larger than the one cited in the Introduction, but
still consistent with lattice QCD calculations, cf.\
the discussion in \cite{petersson}. We choose this
value mainly to maintain consistency with \cite{flow,test2}.},
and $R_A= 1.12\, A^{1/3}$ fm, this leads to
$R/R_A \simeq 0.334\, (\epsilon_0/T_c\, s_c)^{-1/2}$.
Fig.\ 10 shows the results of Fig.\ 8 with the $t_{life}/R$--scale
converted according to this formula. One observes that the lifetime in
units of the radius $R_A$ of the nucleus
does not grow as rapidly with the initial
energy density as in Fig.\ 8, due to the fact that the system's initial size
is shrinking with the CM--frame Lorentz--gamma
(leading to the factor $\epsilon_0^{-1/2}$ in (\ref{ed})).
This suppression of the lifetime at high $\epsilon_0$ has the interesting
effect that existing maxima in the lifetime of different temperatures as
observed in Fig.\ 8 are {\em enhanced} in Fig.\ 10. In the case
$\Delta T=0.1\, T_c$, even new maxima are {\em created\/}
for $T=0.7$ and $0.9\, T_c$. The reduction of the lifetimes of
matter with temperatures near $T_c$ for increasing
$\Delta T$ is, of course, not influenced by the rescaling
of the $t_{life}/R$--axis.

On the other hand, in the ideal gas case, Fig.\ 10 (d), the
rescaling of the $t_{life}/R$--axis creates a {\em minimum}.
This can be understood as follows.
For large $\epsilon_0$ we may employ the asymptotic
formula (\ref{t2}) and note that $R \sim \epsilon_0^{-1/2}$
according to (\ref{ed}), which
results in $t_{life} \sim \epsilon_0^{1/4}$, i.e., a weak increase
with $\epsilon_0$ as observed in Fig.\ 10 (d). For small $\epsilon_0$,
we employ (\ref{t1}) with $T \simeq T_0$, which results in
$t_{life} \sim R \sim \epsilon_0^{-1/2}$, i.e., a decrease with
$\epsilon_0$. The two different forms give rise to the
observed minimum at a certain value of $\epsilon_0$ which increases
for increasing $T/T_c$, since the asymptotic behaviour $T_0 \gg T$
is then realized only for correspondingly larger values of
$\epsilon_0 \sim T_0^4$.
The qualitative difference in the behaviour seen in Fig.\ 10 (c) as
compared to Fig.\ 10 (d) contrasts sharply the similarity of
Figs.\ 8 (c) and (d). The maxima in Fig.\ 10 (c) and
the minima in Fig.\ 10 (d) are, however, not
prominent enough to be of relevance for quark--gluon--plasma
signatures. On the other hand, the huge quantitative
difference in the lifetimes (a factor of $\sim 10$), which is, of course,
still present, could have consequences for observables (see next section).

For the second scenario, we employ the one--dimensional
shock model \cite{schneider,test2}. Since we presently consider
baryon-free matter only, we model the ``ground state'' of the
colliding nuclei as a (massless) pion gas at $T\simeq 143$ MeV, i.e.,
with pressure $p_i = p_c/2$. The energy density $\epsilon_0$
of the hot, compressed state is determined
by solving the integral conservation laws for energy and momentum
across the shock \cite{LL}.
This gives also values for the shock velocity $v_{sh}'$ in the
rest frame of incoming matter
\begin{equation} \label{vsh}
v_{sh}' = \left( \frac{p_0-p_i}{\epsilon_0 - \epsilon_i}~
\frac{\epsilon_0 + p_i}{\epsilon_i + p_0} \right)^{1/2}~,
\end{equation}
and the CM velocity $v_{CM}$ of matter
required to reach that particular energy density,
\begin{equation} \label{vcm}
v_{CM} = \left( \frac{p_0-p_i}{\epsilon_0 + p_i}~\frac{\epsilon_0 -
\epsilon_i}{\epsilon_i + p_0} \right)^{1/2}~.
\end{equation}
The CM time when the complete ``nucleus'' is compressed (or the
shock front has traversed the incoming nucleus, respectively) is
calculated as \cite{bugaev,test2}
\begin{equation}
\frac{t_F}{R_A} = \frac{(1-v_{sh}'v_{CM})\gamma_{CM}}{v_{sh}'}~.
\end{equation}
This time is (up to a factor $c=1$) identical to the
initial ``radius'' $R$ of the system. With equations
(\ref{vsh}) and (\ref{vcm}) one can thus relate $R$ to
$\epsilon_0$, similar as in (\ref{ed}). One has,
however, to consider that a
single shock is not the hydrodynamically stable solution
in thermodynamically anomalous matter \cite{test2}.
The correct solution is a sequence of single shocks and
simple compression waves and, as a consequence, the initial
energy density of the system is no longer constant over
its extension $2\, R$. For our purpose
of obtaining a simple estimate of
$R$ as a function of $\epsilon_0$ we will neglect these details
and use a linear interpolation in the
respective region of $\epsilon_0$--values. The resulting initial
size $R$ of the system as a function of $\epsilon_0$ is shown
in Fig.\ 11.
With this $R(\epsilon_0)$ we then convert
the $t_{life}/R$--scale in Fig.\ 8. The results are depicted in
Fig.\ 12 and are qualitatively similar to Fig.\ 10. The
rapid decrease of $R$ above $\epsilon_0 \simeq \epsilon_Q$ for
$\Delta T= 0$ and $0.01\, T_c$ strongly enhances the maximum in the
lifetime of matter with temperatures near $T_c$ and even
creates a maximum in the lifetime of matter with cooler temperature,
cf.\ Figs.\ 12 (a,b).
The maxima observed in Fig.\ 10 (c) are now also more pronounced, cf.\
Fig.\ 12 (c), while Fig.\ 12 (d) still exhibits minima as in the
corresponding Fig.\ 10 (d).
The one--dimensional shock model thus proves to be the most optimistic
scenario for quark--gluon--plasma signatures that require the
existence of a distinguished
maximum in the lifetime of hot matter at the softest point of
the equation of state, like the proposed enhanced emission of
electromagnetic probes \cite{hung}.

We conclude that the lifetime of matter with temperatures
near $T_c$ is rather sensitive to the choice of
$\Delta T$ and decreases rapidly for increasing $\Delta T$
for initial energy densities around the softest point of the
equation of state. On the other hand, regions of lower
temperature ($T \leq 0.7\, T_c$, cf.\ Fig.\ 9) are less sensitive
to variations in $\Delta T$ and even survive longer with
increasing $\Delta T$. Both phenomena have been explained in terms
of the change in the ``stiffness'' of the equation of state
when varying $\Delta T$. For a fixed initial system size
(independent of $\epsilon_0$), increasing $\Delta T$ eliminates
the maxima in the lifetime near the softest point of the equation
of state (see Fig.\ 8 (c)). On the other hand,
if one accounts for the Lorentz contraction of
the initial size of the system, then such maxima in the
lifetime persist even for finite $\Delta T$.
This is in contrast to the case without a phase transition, where
the lifetimes rather exhibit minima for the corresponding values of
$\epsilon_0$ (Figs.\ 10, 12 (d)).
The lifetimes are also much longer (a factor of
$\sim 10$) in case a rapid change in the degrees of freedom occurs.
Experimental implications will be discussed in the next section.

\section{Summary and Conclusions}

In this work we have studied the one--dimensional hydrodynamic
expansion of a finite system with an equation of state
featuring a transition between (baryon-free) hadronic
matter (with 3 massless degrees of freedom) and a
quark--gluon plasma (with 37 massless degrees of freedom).
The system's evolution was studied as a function of its initial
energy density $\epsilon_0$ (which was assumed to be homogeneous)
and the width $\Delta T$ of the phase transition region
in the equation of state. For $\Delta T < \Delta T^*$, where
$\Delta T^* \simeq 0.07676\, T_c$, the equation of state has
regions where matter becomes thermodynamically anomalous.
Thus, rarefaction shock waves appear in the
one--dimensional expansion, in contrast to the case
$\Delta T \geq \Delta T^*$, where matter is thermodynamically
normal for all energy densities and (overlapping) simple waves form
the hydrodynamically stable expansion solution.

Temperature profiles are not very sensitive to
$\Delta T$ for initial energy densities well above and below
the softest point of the equation of state, in agreement with
the results of \cite{blaizot}.
Variations of $\Delta T$ have, however, an appreciable influence
on the temperature profiles for $\epsilon_0$ near the
softest point. In particular, rarefaction shocks are much weaker for
finite $\Delta T$ than in the case $\Delta T=0$ and transform into
``bumps'' resembling smeared shocks in the thermodynamically normal case
$\Delta T \geq \Delta T^*$. The shock strength is
much smaller because, as shown via eq.\ (\ref{ecrit}),
the region where matter is thermodynamically anomalous shrinks
abruptly when $\Delta T$ becomes finite.
The ``bump'' structures in the thermodynamically
normal case appear due to the variation of the velocity of sound
in the transition region. In general, the
equation of state at high energy densities becomes increasingly stiffer for
increasing $\Delta T$, which leads to more rapid expansion and
cooling of the hot parts of the system. On the other hand,
it becomes softer at smaller energy densities and thus
the cooler parts of the system expand less rapidly and
keep their temperature longer.

The above results were obtained assuming
one--dimensional geometry and neglecting
the finite (net) baryon density in the stopping regime. Nevertheless,
we consider possible experimental consequences that may apply also
in more realistic situations.
The most important consequence from the experimental point of view
is that the lifetimes of matter with temperatures near $T_c$ decrease
rapidly with increasing $\Delta T$,
if $\epsilon_0$ is in the vicinity of the softest point of the
equation of state. This will have an effect on the electromagnetic
signature proposed in \cite{hung}. For instance,
the rate (per unit volume) of direct thermal photons with
energy $E$ emitted from quark--gluon matter (at rest)
is given by \cite{kapusta}
\begin{equation}
E \frac{{\rm d} R^{\gamma}}{{\rm d}^3 {\rm k}} =
\frac{5 \alpha \alpha_S}{18 \pi^2}~T^2
e^{-E/T} \ln \left( \frac{2.912 E}{g^2 T} + 1 \right)~.
\end{equation}
For a fixed photon energy of $E = 1$ GeV, a reduction of
the temperature by $\sim 10 - 20 \%$, as seen in Fig.\ 9 when
$\Delta T$ increases from $0$ to $0.1\, T_c$, would
reduce the number of photons by a factor of $\sim 2 - 4$ (for $T_c=170$
MeV, $\alpha_S=0.4$). In addition, the rate integrated over
space--time will be smaller by an even larger factor, because
the space--time volume of hot matter also shrinks for increasing
$\Delta T$, cf.\ Figs.\ 3 -- 7 (g,h,i). This makes a direct
thermal electromagnetic signal of the quark--gluon plasma much
more difficult to disentangle from background sources.
On the other hand, after taking into account the
expected Lorentz contraction of the initial size of the system,
{\em all\/} lifetimes assume a maximum as a
function of $\epsilon_0$ at the softest point of the equation of state,
cf.\ Figs.\ 11, 12, independent of $\Delta T$. The detection
of this feature is, however, complicated by the broadening
of this maximum with increasing $\Delta T$.

The generic feature that remains {\em independent} of
$\Delta T$ is that the lifetimes of matter of any temperature
are larger by a factor of $\sim 10$ as compared to the ideal gas case.
In this respect, the interferometric
signal of such long lifetimes associated
with a phase transition to the quark--gluon plasma \cite{pratt}
appears to be less sensitive to $\Delta T$ than
hard electromagnetic probes.

Finally we comment on the signature proposed in Ref.\ \cite{flow},
where it was shown that the excitation function of
the directed transverse flow has a minimum
generated by the softest point of the equation of state, or
in other words, the minimum in the function $p/\epsilon$.
Such a minimum still exists even for finite $\Delta T$,
cf.\ Fig.\ 2 (b). However, the intrinsic dynamical mechanism
for the decrease of the transverse directed flow was the inertness
of the mixed phase that prevented it to expand and deflect
spectator matter. This inertness is weakened at finite $\Delta T$
and hot matter in the phase transition region
expands and cools much more rapidly than in the case
of a strong first order phase transition,
although still much slower than if there were no
transition at all, cf.\ Figs.\ 8 (c) and (d).
Further and more detailed, three--dimensional
studies \cite{future} are required before
a definite conclusion of the effect on the transverse
directed flow can be drawn.
\\ ~~ \\
\noindent
{\bf Acknowledgments}
\\ ~~ \\
We thank A.\ Dumitru for discussions and
the Nuclear Theory Group at Brookhaven
National Laboratory for hosting a summer workshop, where part
of this work was done.

\newpage
\noindent
{\bf Figure Captions:}
\\ ~~ \\
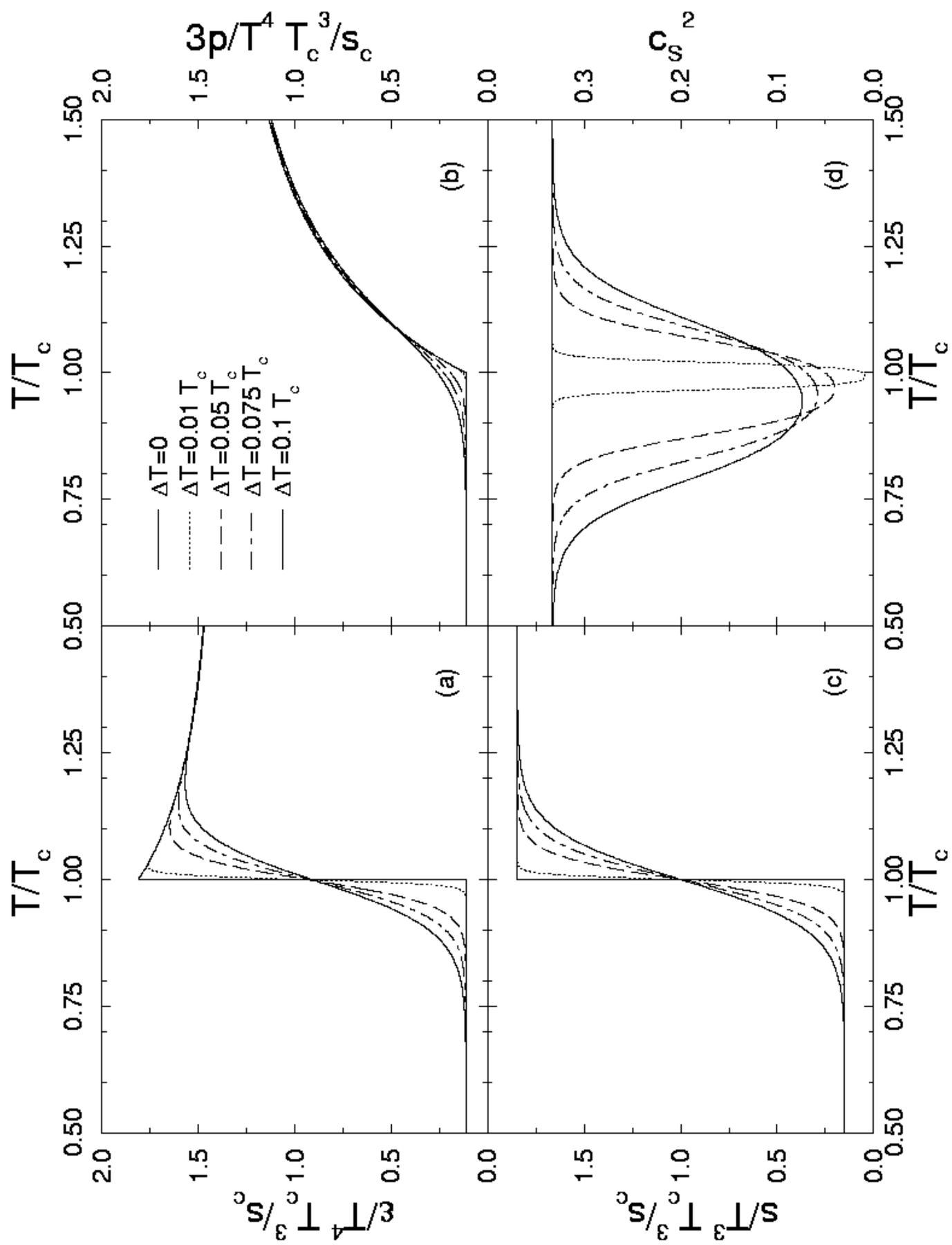
{\bf Fig.\ 1:} (a) the energy density (in units of $T_c\, s_c$)
divided by $(T/T_c)^4$, (b) three times the pressure (in units of
$T_c\, s_c$) divided by $(T/T_c)^4$, (c) the entropy density
(in units of $s_c$) divided by $(T/T_c)^3$, and (d) the square
of the velocity of sound as functions of $T/T_c$. The solid line
with the discontinuity at $T=T_c$ in (a,c) corresponds to
$\Delta T=0$. In (b), this curve has a kink at $T=T_c$, and in
(d) it is the flat line at $c_s^2=1/3$ (the point $T=T_c$
is an exception, there $c_s^2=0$).
The other curves correspond to $\Delta T=0.01\, T_c$ (dotted),
$\Delta T=0.05\, T_c$ (dashed), $\Delta T=0.075\, T_c$
(dash-dotted), and $\Delta T=0.1\, T_c$ (solid).
\\ ~~ \\
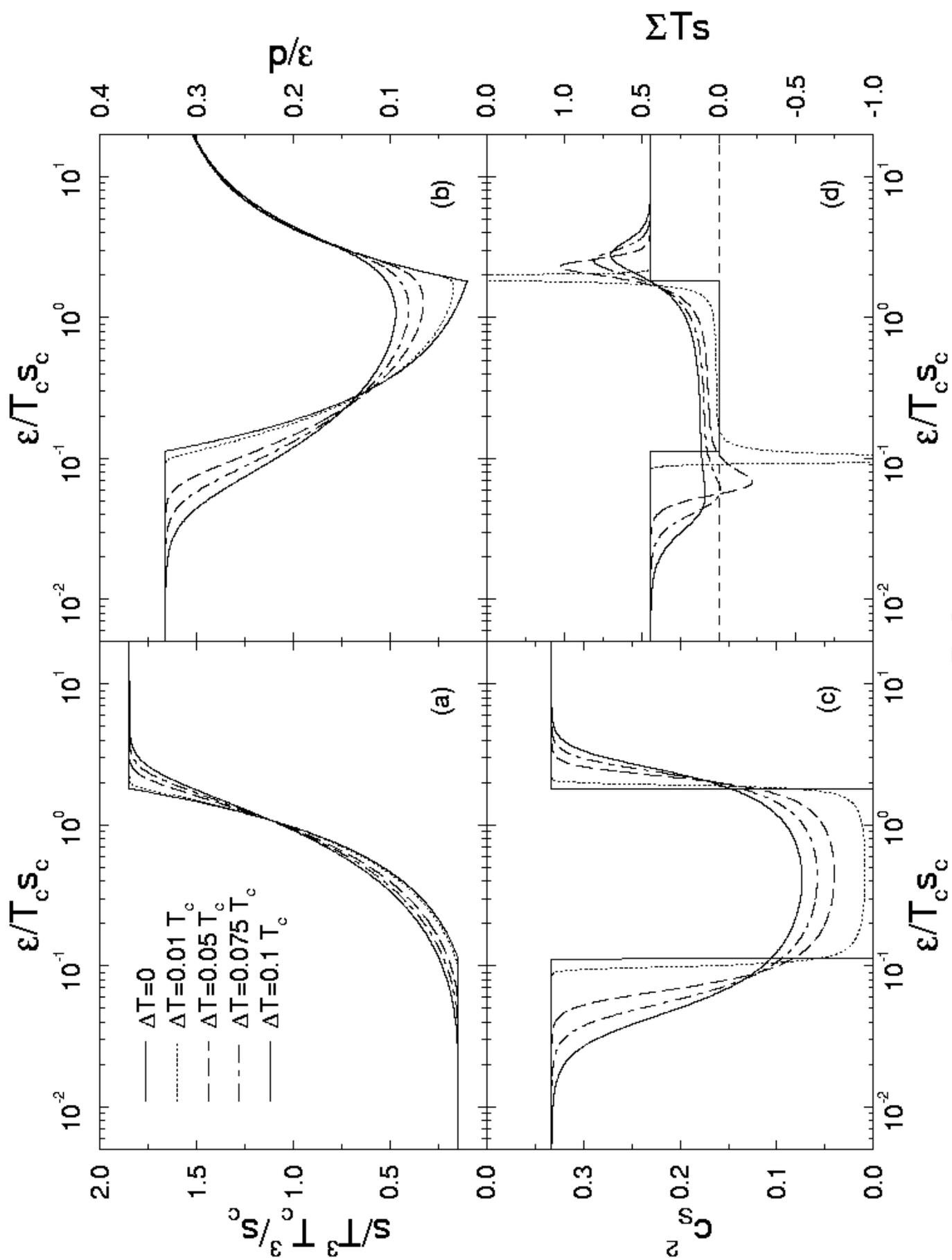
{\bf Fig.\ 2:} (a) entropy density (in units of $s_c$),
divided by $(T/T_c)^3$, (b) $p/\epsilon$, (c) the velocity
of sound squared, and (d) $\Sigma Ts$ as functions of
energy density (in units of $T_c\, s_c$). The different
curves correspond to those in Fig.\ 1.
\\ ~~ \\
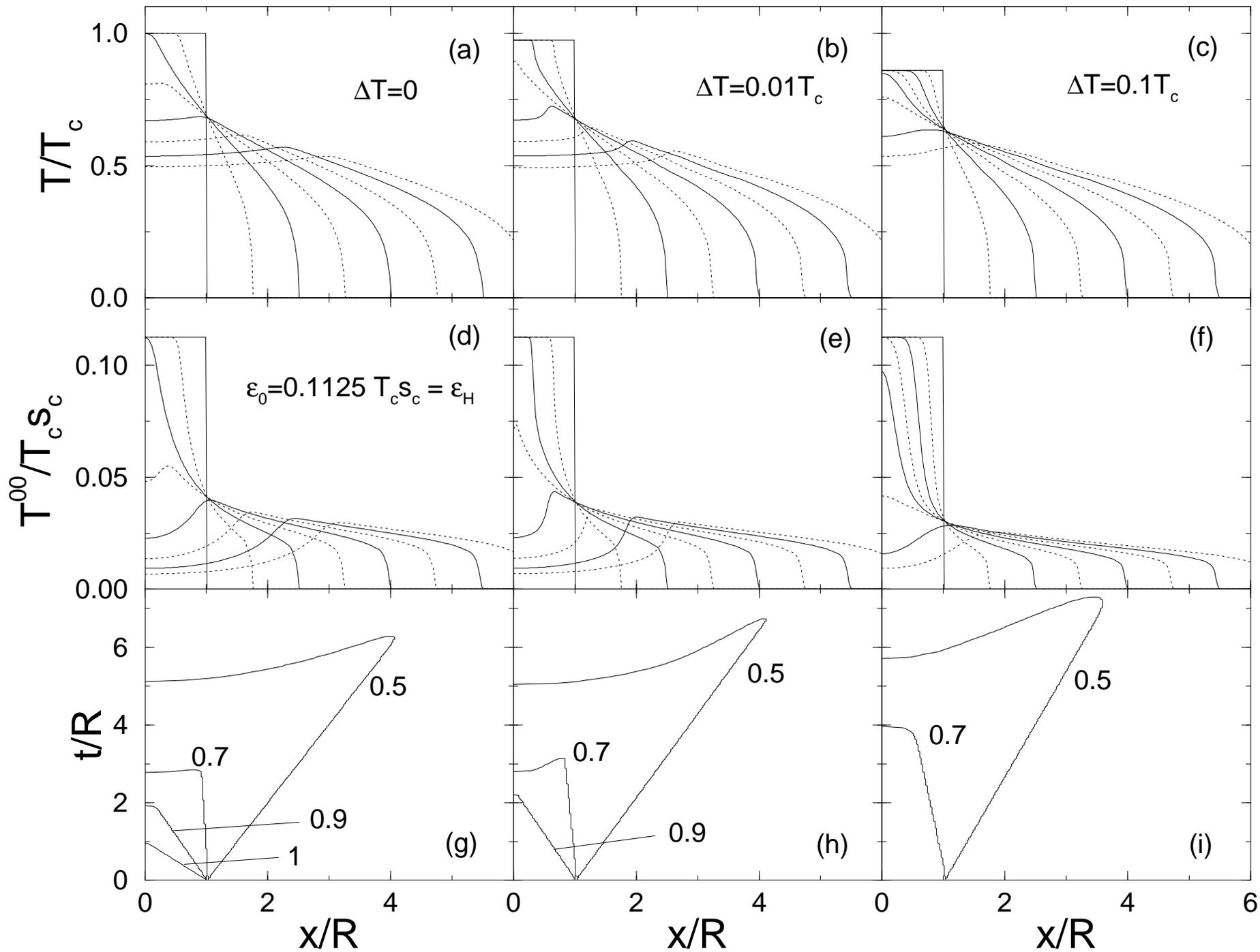
{\bf Fig.\ 3:} (a,b,c) temperature profiles and (d,e,f)
center-of-momentum frame energy density profiles for times $t=0.75\,
n \lambda R,\, n=0,1,...,7$ and an initial energy
density of $\epsilon_0 = 3\, T_c\, s_c/2(r+1) =
0.1125\, T_c\, s_c \equiv \epsilon_H$. The profiles
are alternatingly shown as full and dotted lines in order to better
distinguish them. (g,h,i) show isotherms in the space--time diagram.
The isotherms are labelled with the
corresponding temperatures in units of $T_c$. Figs.\
(a,d,g) are for $\Delta T=0$, (b,e,h) for
$\Delta T=0.01\, T_c$, and (c,f,i) for $\Delta T=0.1\, T_c$.
\\ ~~ \\
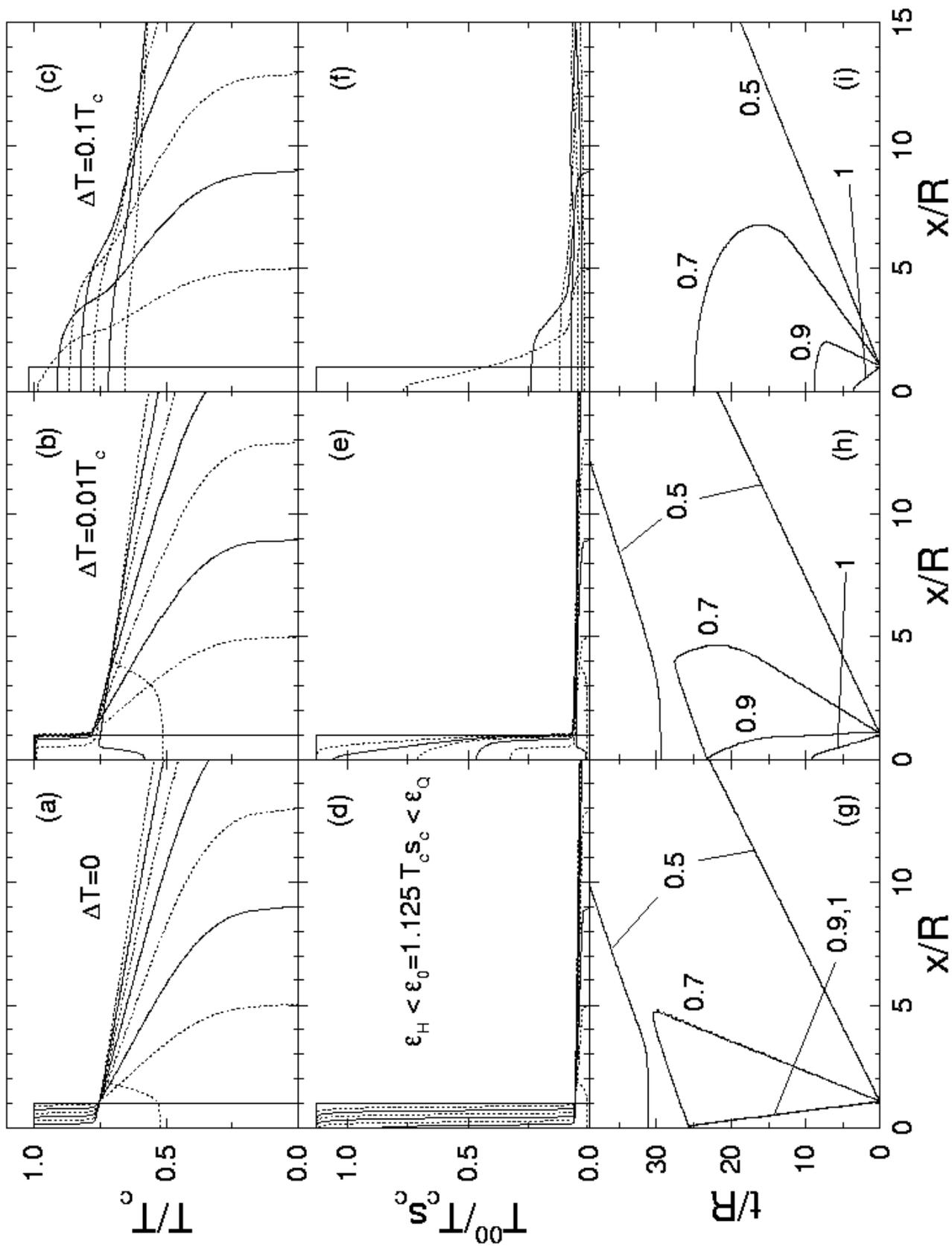
{\bf Fig.\ 4:} Same as in Fig.\ 3, for
$\epsilon_0 = 1.125\, T_c\, s_c$, i.e., $\epsilon_H <
\epsilon_0 < \epsilon_Q = (4r-1)\, T_c\, s_c/2(r+1)$.
Profiles in (a--f) are for times $t=4\, n \lambda R,\, n=0,1,...,7$.
\\ ~~ \\
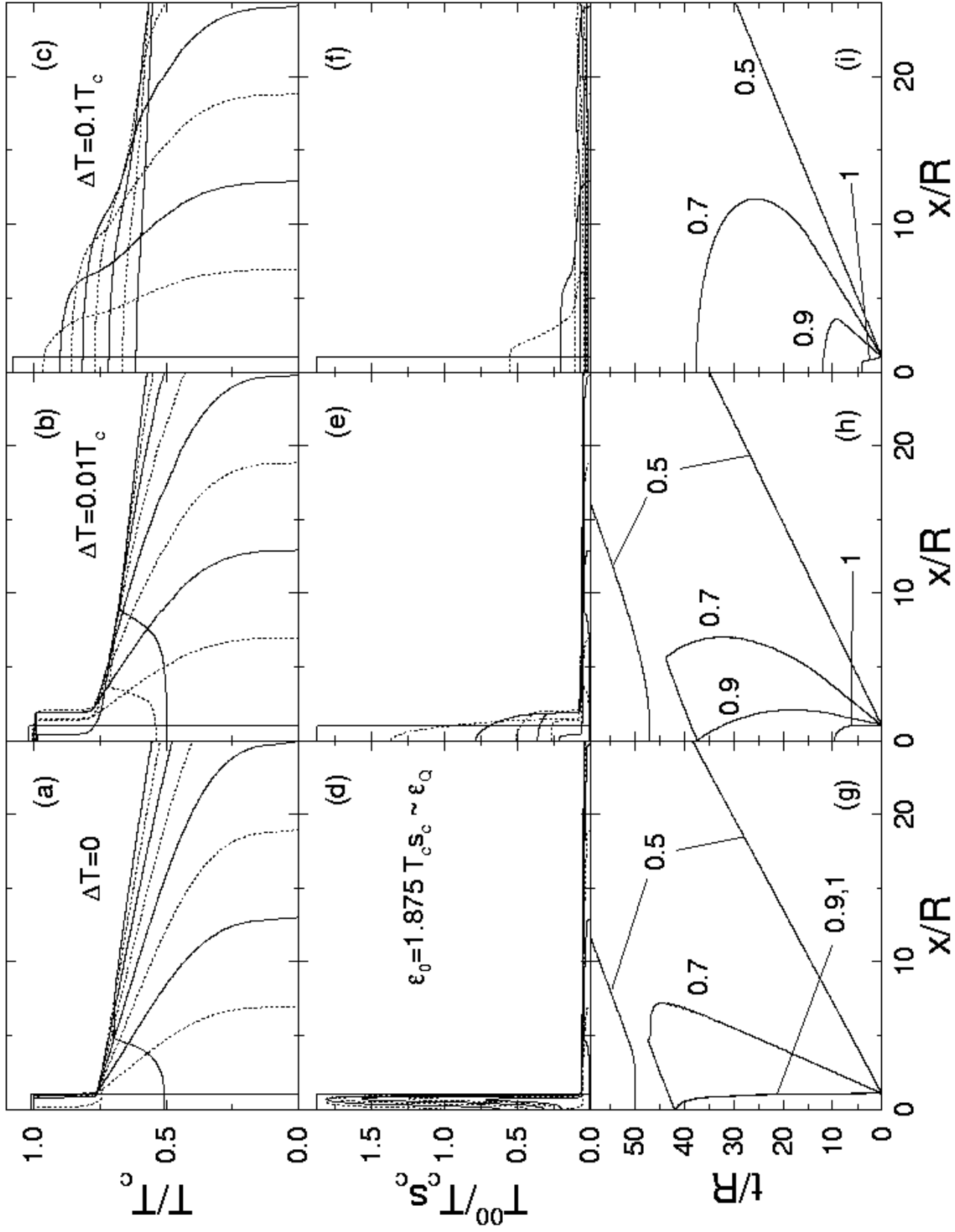
{\bf Fig.\ 5:} Same as in Fig.\ 3, for
$\epsilon_0 = 1.875\, T_c\, s_c \sim \epsilon_Q$.
Profiles in (a--f) are for times $t=6\, n \lambda
R,\, n=0,1,...,8$.
\\ ~~ \\
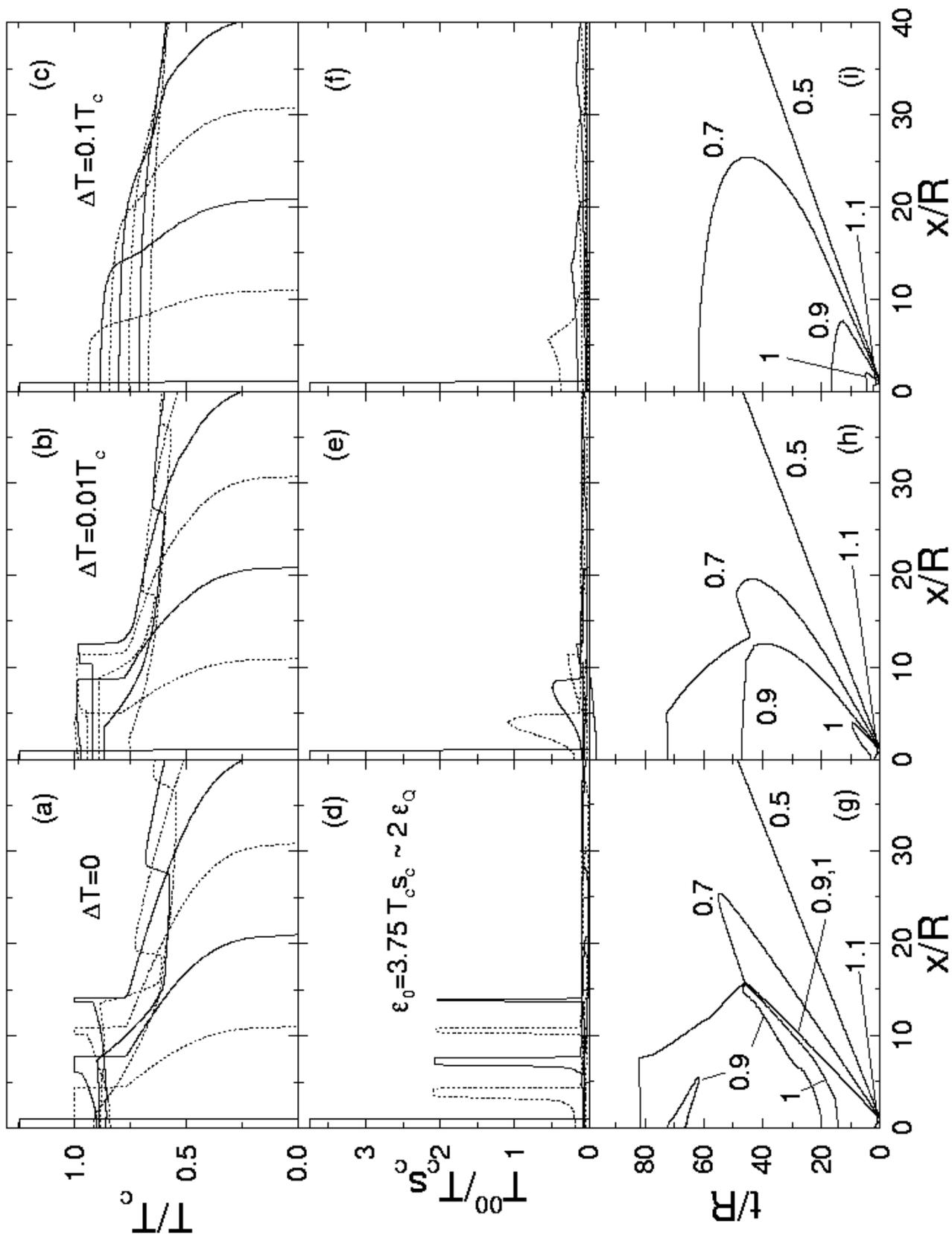
{\bf Fig.\ 6:} Same as in Fig.\ 3, for $\epsilon_0 = 3.75\, T_c\, s_c
\sim 2\, \epsilon_Q$. Profiles in (a--f) are for times $t=10\, n \lambda
R,\, n=0,1,...,7$.
\\ ~~ \\
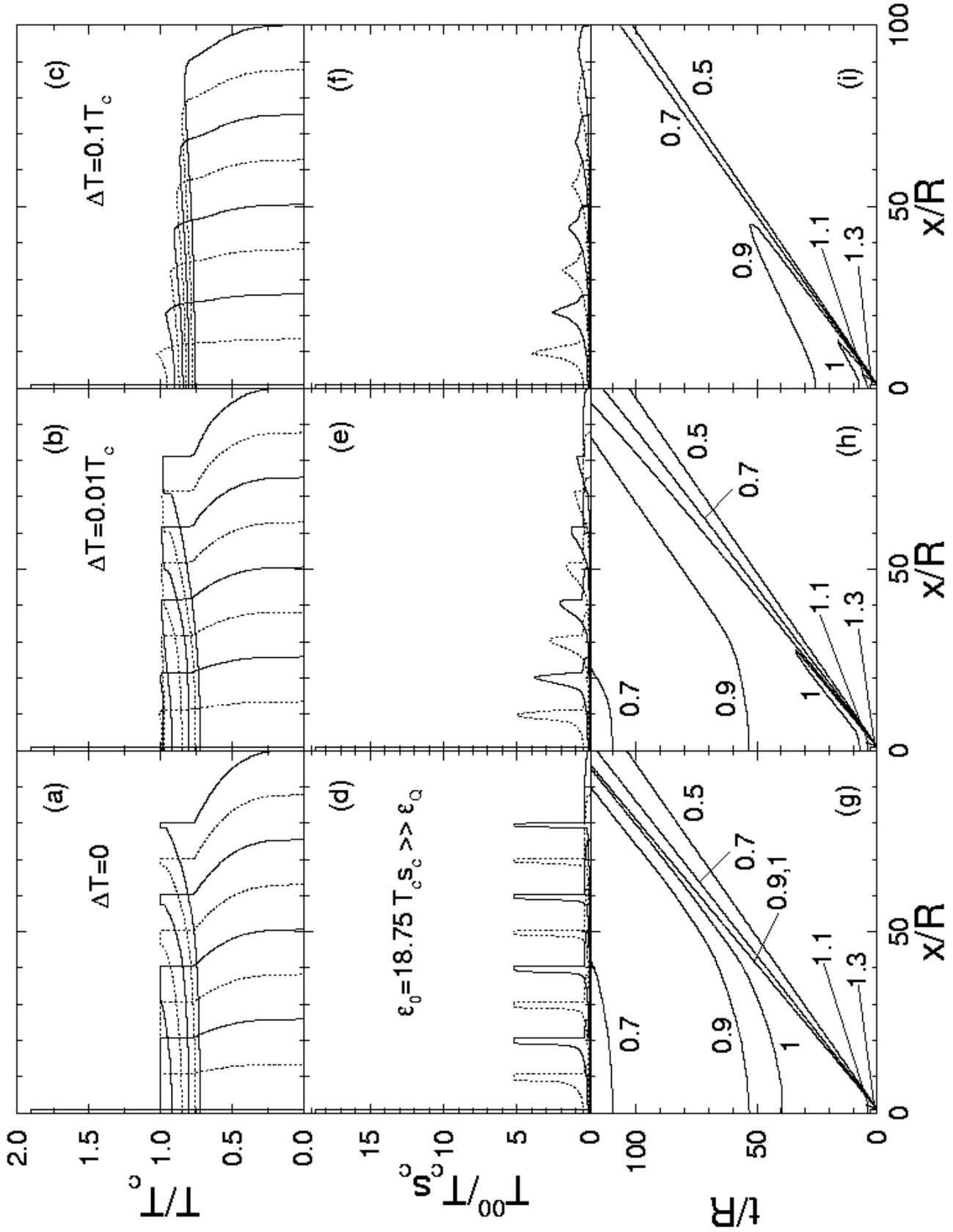
{\bf Fig.\ 7:} Same as in Fig.\ 3, for $\epsilon_0 = 18.75\, T_c\, s_c
\gg \epsilon_Q$. Profiles in (a--f) are for times $t=12.5
\, n \lambda R,\, n=0,1,...,8$.
\\ ~~ \\
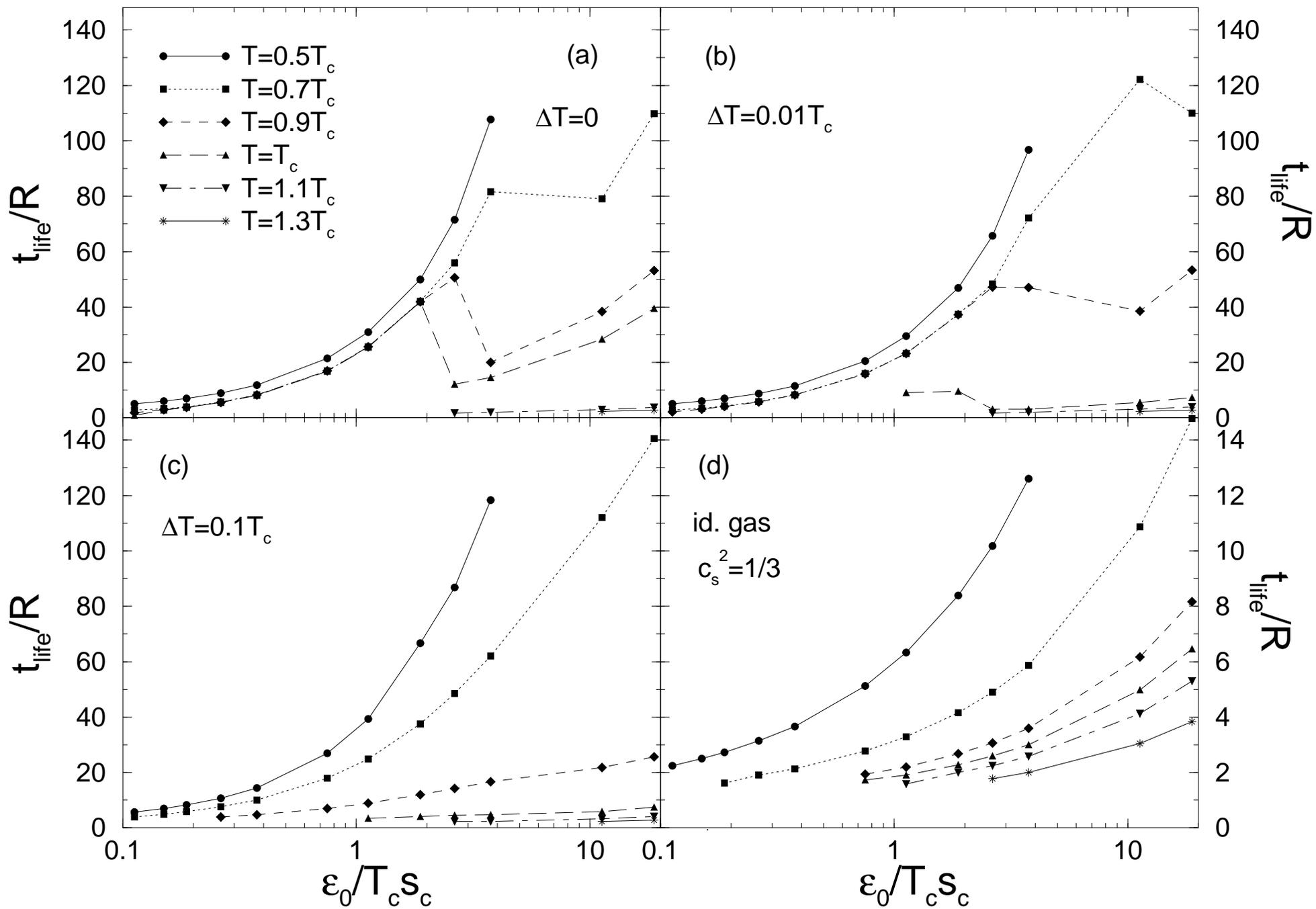
{\bf Fig.\ 8:} The lifetime of differently hot parts of the
system as a function of the initial energy density for (a)
$\Delta T=0$, (b) $\Delta T=0.01\, T_c$, (c)
$\Delta T=0.1\, T_c$, and (d) an ideal gas
equation of state with $c_s^2=1/3$. Full line and circles: lifetime of
matter having temperature $T=0.5\, T_c$, dotted line
and squares: $T=0.7\, T_c$, dashed line and diamonds: $T=0.9\,
T_c$, long dashed line and triangles pointing upwards:
$T=T_c$, dash-dotted line and triangles pointing downwards:
$T=1.1\, T_c$, full line and stars: $T=1.3\, T_c$.
Note the change in the scale of $t_{life}$ in (d).
\\ ~~ \\
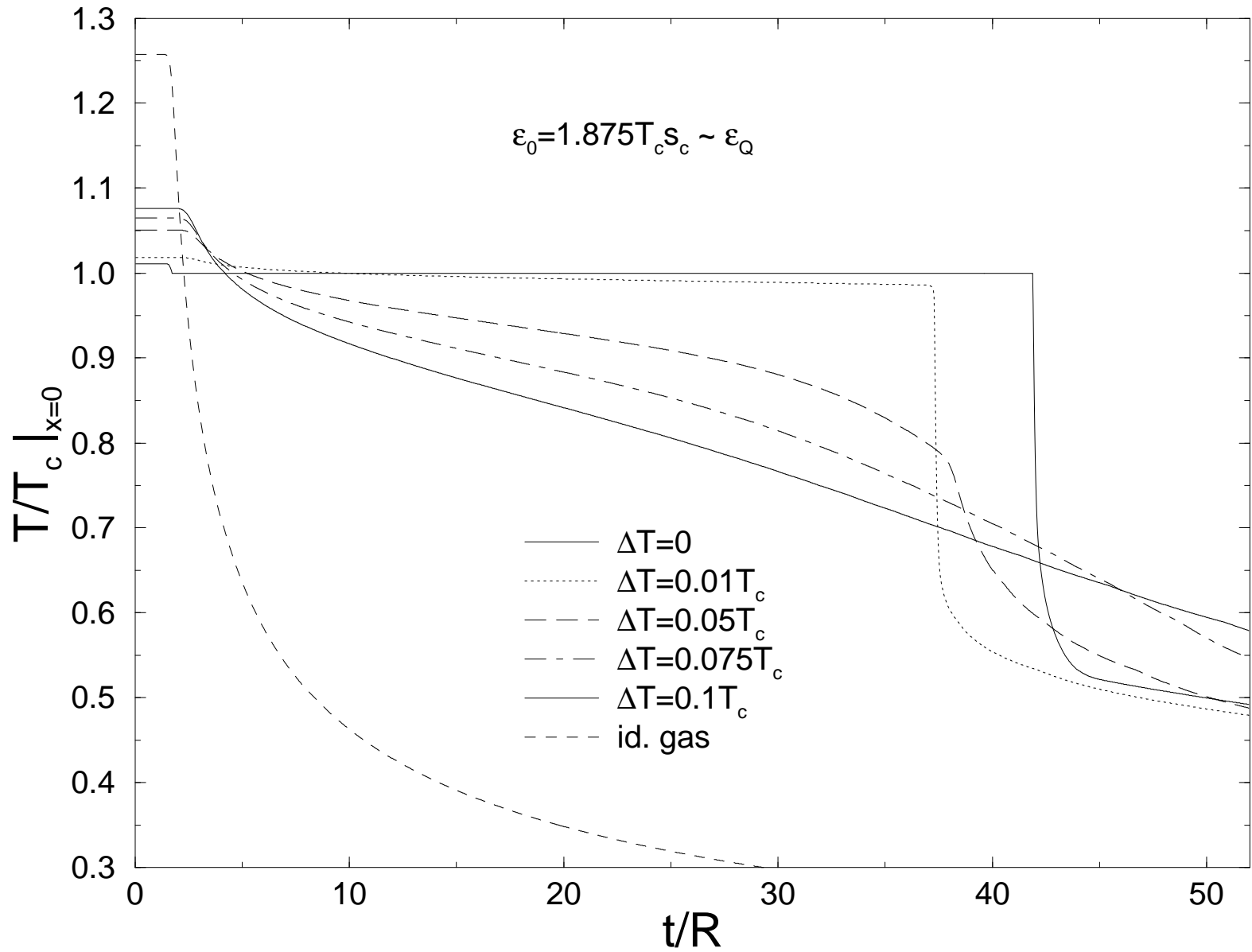
{\bf Fig.\ 9:} Temperature (in units of $T_c$) at $x=0$
as a function of time (in units of $R$)
for $\epsilon_0 = 1.875\, T_c\, s_c \sim \epsilon_Q$.
Lines correspond to $\Delta T$--values as in Figs.\ 1, 2.
The dashed line corresponds to a calculation with an
ideal gas equation of state, $c_s^2 = 1/3$.
\\ ~~ \\
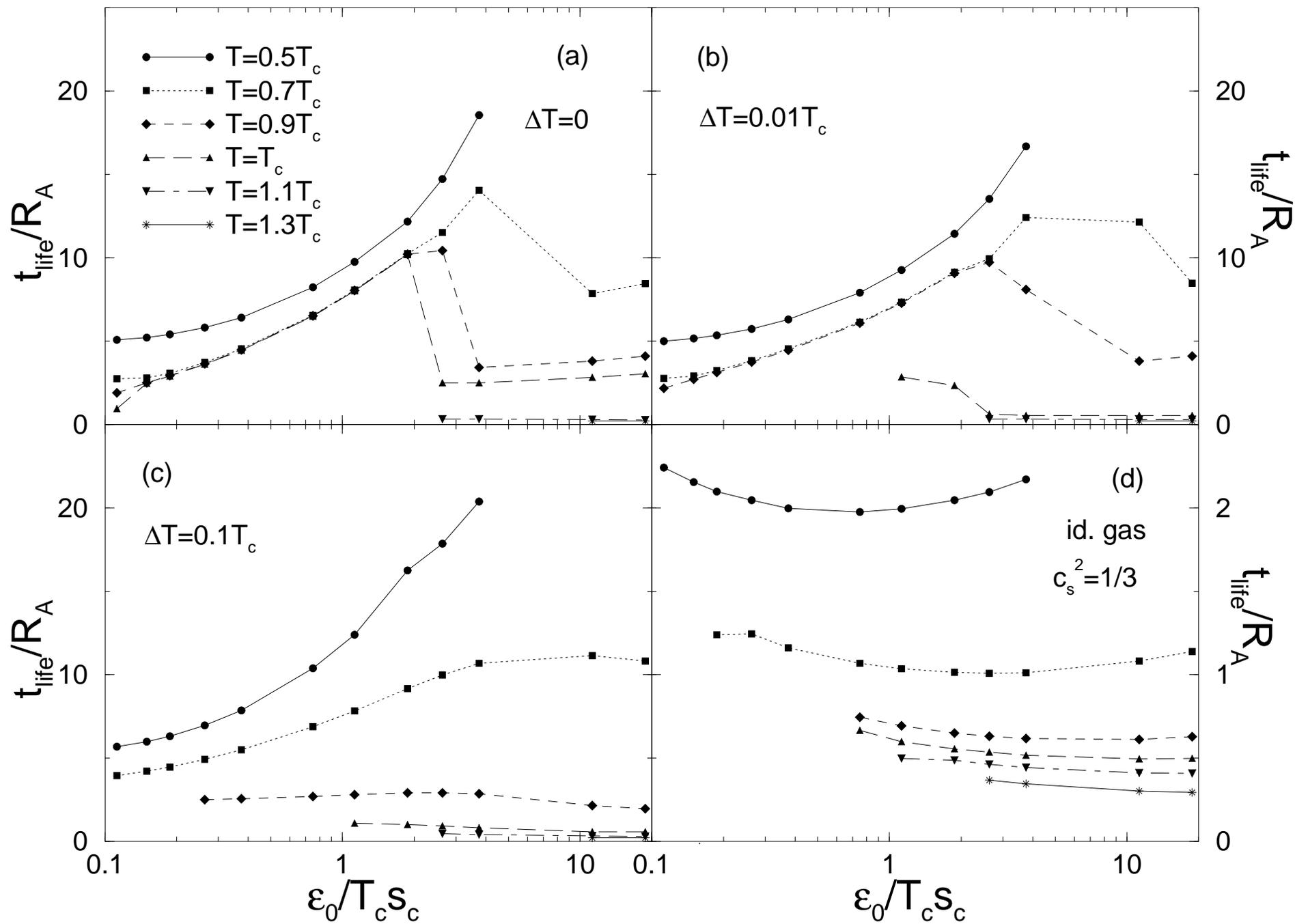
{\bf Fig.\ 10:} The same as in Fig.\ 8, except that $t_{life}/R$
is converted to $t_{life}/R_A$ according to eq.\ (\ref{ed}),
where $R_A$ is the radius of a nucleus at rest.
\\ ~~ \\
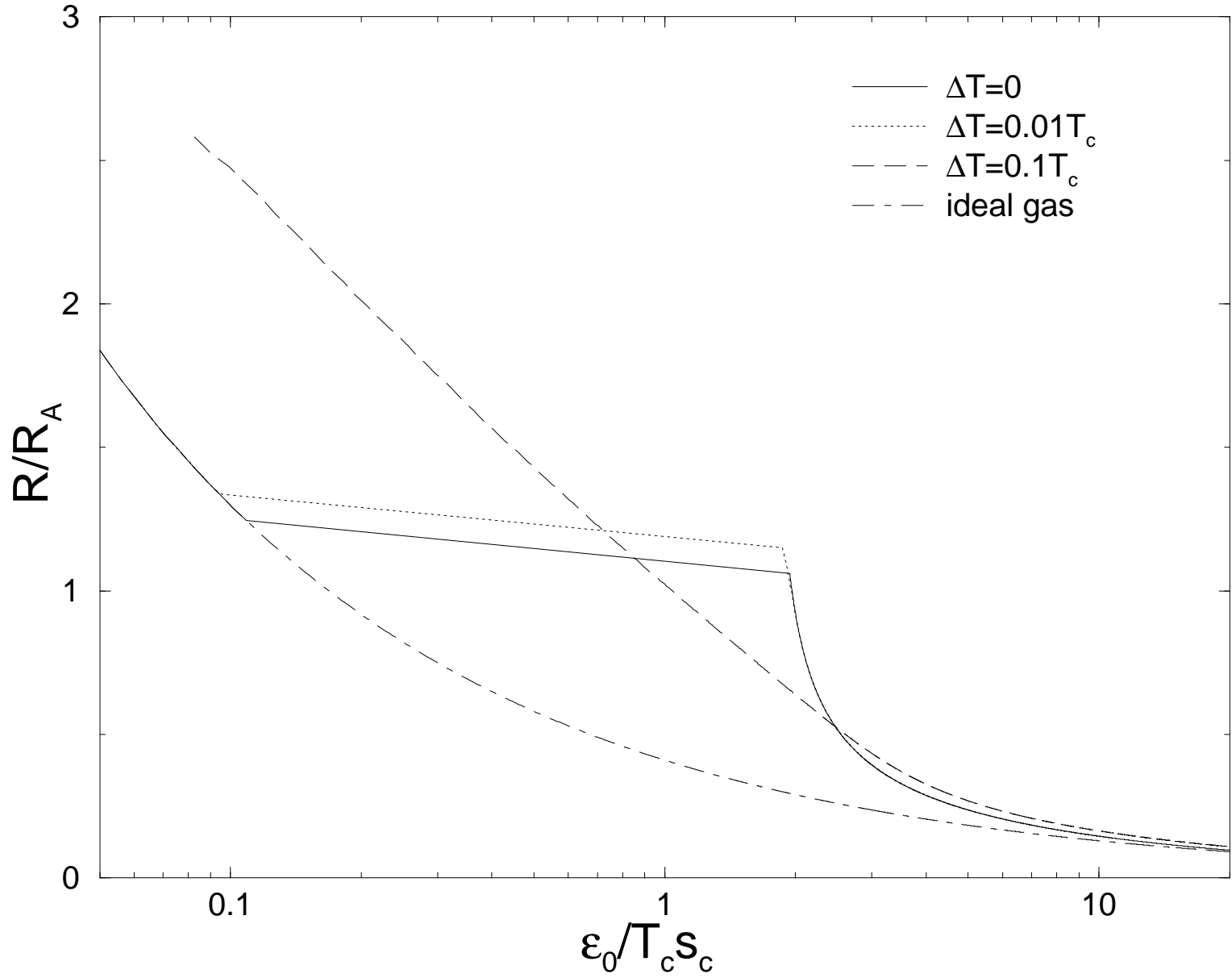
{\bf Fig.\ 11:} The system's longitudinal ``radius'' $R$ in units
of $R_A$ as a function
of the initial energy density in the one--dimensional shock model.
Full line: $\Delta T=0$, dotted line: $\Delta T=0.01\, T_c$,
dashed line: $\Delta T=0.1\, T_c$, dash-dotted line: ideal gas
equation of state. The straight parts of the full and dotted lines
are a linear interpolation between the points where the
hydrodynamically stable compression solution ceases to be a single
shock due to the thermodynamically anomalous behaviour of the
equation of state in the phase transition region.
\\ ~~ \\
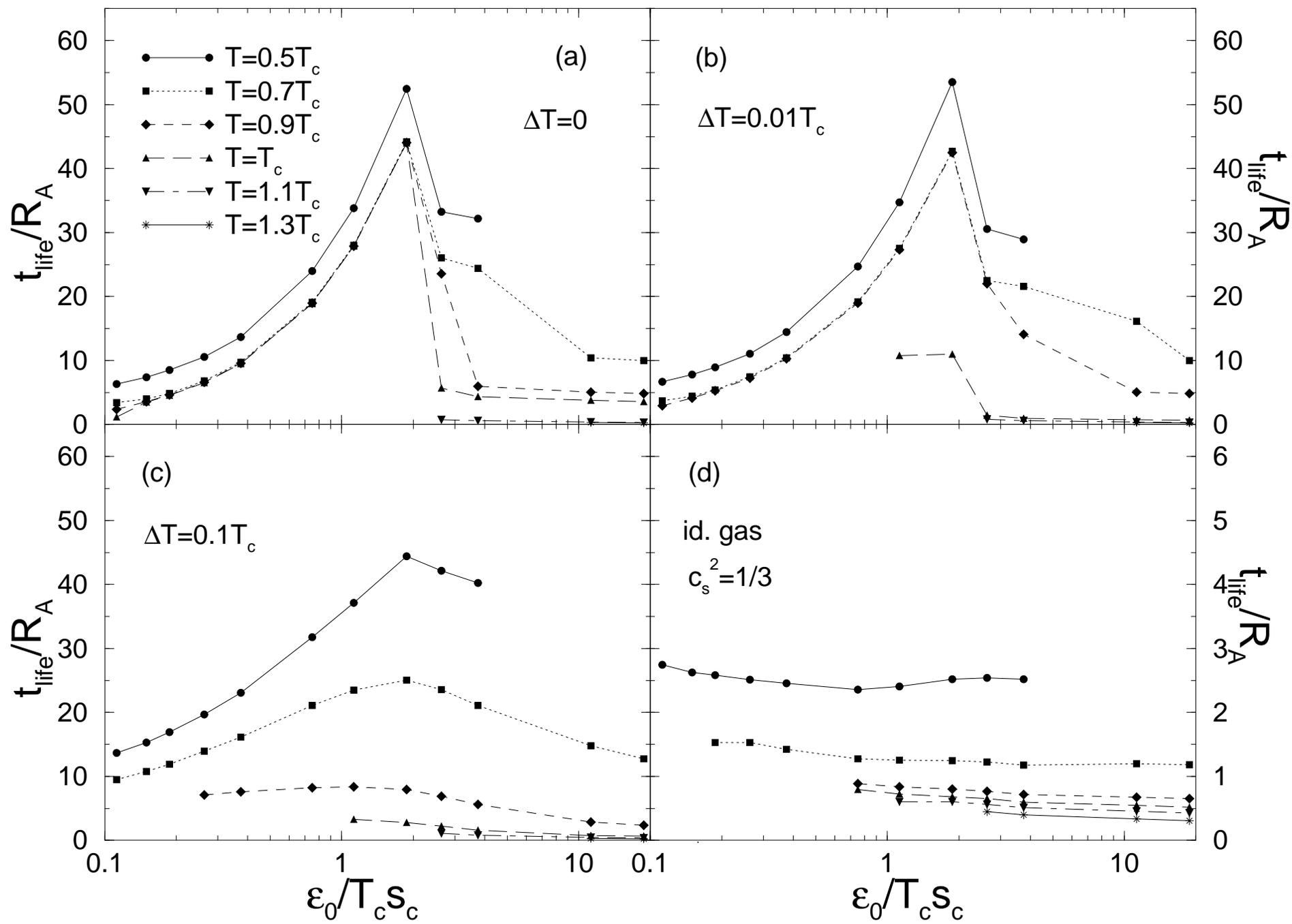
{\bf Fig.\ 12:} The same as in Fig.\ 8, except that $t_{life}/R$
is converted into $t_{life}/R_A$ according to Fig.\ 11.
\end{document}